\documentclass[3p, preprint]{elsarticle}
\usepackage{ecrc}
\usepackage{amssymb}
\usepackage{amsthm}
\usepackage{eqnarray,amsmath}
\usepackage{setspace}
\usepackage{url} 
\usepackage{hyperref}
\usepackage{xcolor}
\hypersetup{colorlinks,citecolor=green,linkcolor=red,urlcolor=blue}

\usepackage{ulem}
\newcommand{\GeV}{\textrm{GeV}}

\runauth{M. Guzzi, T. Hobbs, K. Xie, et al.}
\journal{Physics Letters B}
\jid{elsarticle}
\jnltitlelogo{Physics Letters B}
\firstpage{1}
\volume{00}

\begin{document}

\begin{frontmatter}

\title{The persistent nonperturbative charm enigma}

\author{Marco Guzzi}
\address{Department of Physics, Kennesaw State University, Kennesaw, GA 30144, USA}

\author{T.~J.~Hobbs \corref{cor1}}
\address{High Energy Physics Division, Argonne National Laboratory, Argonne, IL 60439, USA}
\cortext[cor1]{Corresponding author: \href{mailto:tim@anl.gov}{tim@anl.gov}}

\author{Keping Xie}
\address{Pittsburgh Particle Physics,
  Astrophysics, and Cosmology Center,\\ Department of Physics and
  Astronomy, University of Pittsburgh, Pittsburgh, PA 15260,
  USA\looseness=-1}

\author{Joey Huston}
\address{Department of Physics and Astronomy, Michigan State University, East Lansing, MI 48824, USA}

\author{Pavel Nadolsky}
\address{Department of Physics, Southern Methodist University,
  Dallas, TX 75275-0181, USA}
\address{Fermi National Accelerator Laboratory, Batavia, IL 60510, USA}

\author{C.-P. Yuan}
\address{Department of Physics and Astronomy, Michigan State
  University, East Lansing, MI 48824, USA\looseness=-1}

\begin{abstract}
The question of the existence and possible magnitude of nonperturbative (often called ``intrinsic'') charm in the proton has
long confounded attempts to cleanly isolate such a contribution in global analyses of high-energy experiments.
In this letter, we show that the available (non)perturbative QCD theory and hadronic data have still not developed to a sufficient level to clearly resolve this problem.
We highlight a number of challenging aspects that must be confronted in extracting nonperturbative charm
in PDF fits, and in so doing, present an updated next-to-next-to-leading order CT analysis of fitted charm, CT18 FC, which we also compare to recent studies.
We outline the theory developments and future data needed to make progress on this subject.
\end{abstract}

\dochead{\begin{flushright} \small Preprint ANL-179603, FERMILAB-PUB-22-786-T, MSUHEP-22-035, SMU-HEP-22-10 \end{flushright}}
\date{\today}

\begin{keyword}
Quantum Chromodynamics \sep parton distribution functions \sep collider phenomenology

\end{keyword}
\end{frontmatter}

\clearpage \newpage

\paragraph{\bf 1.~Introduction}
\hypertarget{sec:intro}
The possibility that the nucleon might harbor a small but nonzero nonperturbative charm component~\cite{Brodsky:1980pb}
was identified soon after the establishment of QCD as the microscopic theory of the strong interaction. This notion, which has variously been called ``intrinsic,'' ``nonperturbative,'' or ``fitted'' charm, has been challenging to formalize in an unambiguous fashion based
on rigorous QCD. Despite this, multiple efforts have attempted to isolate a nonperturbative charm
PDF through a global QCD analysis of the available hadronic data, a class of approaches we designate {\it fitted charm} (FC).
These works often assume specific nonperturbative shapes for the FC PDF at the evolution starting-scale, $Q_0$, based on QCD-inspired {\it intrinsic charm} (IC) models,
producing inconclusive results regarding the possible
magnitude. The overall magnitude and $c$, $\bar{c}$ asymmetry induced by nonperturbative charm is often quantified in the moments
of the charm PDF~\cite{Pumplin:2007wg,Jimenez-Delgado:2014zga,Hou:2017khm},
$\langle x^n \rangle_{c^\pm}(Q)\! \equiv\! \int^1_0 dx\, x^n\, (c \pm \bar{c})[x,Q]$, where the $n\!=\!1$ moment of
the ``+'' combination yields the total proton momentum carried by the charm (anti)quark.

On the basis of the momentum fraction or related quantities, various definite but conflicting claims have been made
in the literature with respect to nonperturbative charm.
In this Letter, we highlight lingering challenges by reviewing the current status of FC on the basis of recent QCD theory, and by presenting a global QCD analysis within the next-to-next-to-leading order (NNLO) CTEQ-TEA (CT) framework that updates in-depth studies of the underlying theory and phenomenology published a few years ago \cite{Jimenez-Delgado:2014zga,Hou:2017khm}.
After a discussion of open issues in the status, definition, and implications of nonperturbative charm from the perspective of QCD theory
(\hyperlink{sec2}{Sec.~2}), in \hyperlink{sec3}{Sec.~3} we turn to the posited sensitivity of several experimental measurements to nonperturbative charm.
Following this, in \hyperlink{sec4}{Sec.~4} we introduce a new family of PDFs with FC, CT18 FC, and compare their behavior and agreement with data against analogous findings from other recent studies. We point out that nonperturbative QCD effects may lead to a difference between charm quark and antiquark PDFs at the low-energy scale, as we illustrate through the example of a particular pair of IC models. In \hyperlink{sec5}{Sec.~5}, we contrast against the recent NNPDF analysis \cite{Ball:2022qks}, which reported a $3\sigma$-level
extraction of ``intrinsic charm'' based on an $x$-dependent deviation of a FC PDF from a purely radiatively-generated
charm scenario.
We assess the significance of the suggested evidence in light of subtleties in the analysis methodologies used in FC studies.
Finally, we conclude (\hyperlink{sec6}{Sec.~6}) with recommendations for future studies on nonperturbative charm at the Large Hadron Collider (LHC) and Electron-Ion Collider (EIC). Additional, detailed quantitative results are provided in the Supplementary Discussion (SD) section.


\paragraph{\bf 2.~Nonperturbative charm in QCD}
\hypertarget{sec2}{The} argument for a finite nonperturbative charm component in the proton is grounded in low-energy QCD. Open questions nonetheless remain concerning the rigorous definition, potential process (in)dependence,
and actual magnitude of this contribution \cite{Hou:2017khm}.
An enduring challenge in studies of IC has been the absence
of a universal definition derived formally from QCD and free of ambiguities related
to the interpretation.
Traditionally, IC has been argued~\cite{Brodsky:1980pb,Brodsky:1984nx} to arise from the production of charm quark-antiquark pairs in long-distance
QCD interactions.
Despite its small magnitude, IC may in principle still be discernible from the perturbative (``extrinsic") production of charmed final states through radiation in independent scatterings off initial-state gluons and light (anti)quarks.
As IC processes are thought to generate a nonzero charm distribution at $Q\!\sim\!m_c$, they act over sufficiently soft momenta as to be inherently
nonperturbative; as such, the associated dynamics has been formulated through models which typically couple the proton to 5-quark intermediate
states~\cite{Brodsky:1980pb,Chang:2011vx,Hobbs:2017fom} explicitly containing $c, \bar{c}$. From this simplified picture,
various elaborations are possible, involving, {\it e.g.}, production of intermediate hadronic (meson-baryon) states~\cite{Pumplin:2005yf,Hobbs:2013bia} or multi-quark virtual states with different spin structure~\cite{Hobbs:2017fom}.

Any IC PDF is ultimately a scheme-dependent function, analogous to the $\overline{\mathrm{MS}}$ charm mass. It is not a physical scattering contribution that can be
directly measured and thereby unambiguously discovered.
Borrowing an analogy to nucleon strangeness, it is tempting to freely parametrize and fit $xc^+$ at or near the charm threshold
and identify the resulting FC PDF with IC.
We again stress that the FC PDFs extracted in recent examples of this approach, including CT14 IC \cite{Hou:2017khm},
NNPDF~\cite{Ball:2022qks,NNPDF:2017mvq,Ball:2016neh}, and Ref.~\cite{Jimenez-Delgado:2014zga}, are actually approximations of IC,
due to the possibility that the post-fit parametrization may absorb contributions unrelated to IC.
This is reinforced by the fact that there is no unambiguous mapping between the IC PDFs predicted in nonperturbative models and FC PDFs that might be extracted
in global fits based on QCD factorization;
this ambiguity includes the fact that IC models should apply at an energy scale $Q$ that is indeterminate.
Furthermore, compared to perturbatively-generated charm, the magnitude of FC quantified in this way depends on various theoretical parameters in the PDF analysis, such as the value of the charm-quark mass, $m_c$; QCD coupling strength, $\alpha_s$; and auxiliary energy scales introduced in massive-quark factorization schemes.
In fact, the FC PDF itself is analogous to the fitted-charm mass, whose best-fit value may be affected by corrections leading to deviations from the charm mass in the QCD Lagrangian.
In addition, it is not clear that indications of IC are process-independent, as would be necessary to claim that the IC is a universal component of the proton wave function.
Ref.~\cite{Hou:2017khm} provides a systematic ordering of leading-power and power-suppressed (higher-twist) contributions in DIS to understand the IC component in the framework of QCD factorization, and it illustrates the IC dependence on theory parameters at NNLO. 
Without theory developments to connect FC extracted in PDF analyses to the IC from traditional models, it is impossible
to guarantee that the fitted ``IC'' is in fact a universal contribution to the proton wave function, rather than a process-dependent,
non-leading twist, or other effect that has been spuriously absorbed.

Still, many attempts have been made to constrain IC in QCD global fits, recently including CT14 IC \cite{Hou:2017khm},
NNPDF~\cite{Ball:2022qks}, and Ref.~\cite{Jimenez-Delgado:2014zga}.
While these studies have elucidated many aspects of the IC, inconclusive hints from experiments have not yet reached discovery-level, particularly given the subtleties discussed above.


\paragraph{\bf 3.~Experimental signatures of nonperturbative charm}
\hypertarget{sec3}{Unraveling} a possible nonperturbative charm PDF at high $x$ is empirically challenging, owing both
to its small net momentum fraction, [$\langle x \rangle_\text{FC}\! \equiv\! \langle x \rangle_{c^+}(Q_0)\! \lesssim\! 10^{-2}$], and the difficulty of performing the necessary flavor separation.
Apart from the classical search channel in large-$x$ DIS charm production \cite{EuropeanMuon:1982xfn,Hoffmann:1983ah}, few hadronic data sets have been identified in previous QCD global fits as having clear `smoking gun' sensitivity to nonperturbative charm, though it has been suggested~\cite{Ball:2022qks} that the bulk of recent $pp$ data from the LHC provide sufficient sensitivity to unravel IC.
Many of these hadronic data sets have been independently
investigated by the CT and other fitting groups and jointly explored in the recent PDF4LHC'21 exercise \cite{PDF4LHCWorkingGroup:2022cjn}; for these sets, there is no clear preference for FC in CTEQ-TEA fits; we reconfirm this point in the updated CT refit of FC shown in \hyperlink{sec4}{Sec.~4}.
In the SD section, we provide several plots illustrating the generally marginal pull of the latest
LHC data on the FC PDF; these pulls can in fact become intermixed
with the light-quark sea, obfuscating the relationship with
nonperturbative charm.
In terms of measurements with more direct access, $pp\! \to\! Z\!+\!c$ has been suggested~\cite{Boettcher:2015sqn,Bailas:2015jlc}
as having elevated sensitivity to the high-$x$ charm PDF and possible IC scenarios. Fully leveraging these data
requires that they be consistently treated at the current standard in PDF analyses for high-energy physics (HEP) --- NNLO.
While NNLO QCD predictions have been published for $Z\!+\!b$-jet and $W\!+\!c$-jet production at the LHC recently~\cite{Gauld:2020deh,Czakon:2020coa}, and for $Z+c$ production \cite{Gauld:2023zlv} after the initial release of this article, these have not been practically incorporated into PDF fits. In the $Z+c$ process, final-state parton showering and hadronization introduce a large correction at NLO that dampens the excess at high $p_T$ typically induced by large-$x$ IC; see a quantitative investigation of the impact of parton showers on the sensitivity to various IC models in Sec.~6 of Ref.~\cite{Hou:2017khm}, as well as the very latest work in \cite{Gauld:2023zlv}. 
In Fig.~2 of the SD, we show theory predictions for the 2022 LHCb $\sigma(Zc)/\sigma(Zj)$ ratios~\cite{LHCb:2021stx} based on MCFM
at NLO and several PDF sets with and without FC; these calculations show that inclusion
of FC does little to enhance agreement with the experimental data, while 
significant uncertainties exist in current theory predictions for this process.
Additionally, in the available $Z+c$ measurements, the anti-$k_T$ jet algorithm with charm tagging applied at the experimental level differs from those needed for QCD calculations to be infrared-safe (flavor-$k_T$~\cite{Banfi:2006hf} or flavored anti-$k_T$~\cite{Czakon:2022wam}). This enhances the sensitivity to the details of treatment of charm quarks with low $p_T$ in the sample, {\it e.g.}, to the specific choice of the jet cone size and $p_T^c$ cut of a few GeV. In the $b$-jet measurement, a related uncertainty is estimated to be about 10\% ~\cite{Czakon:2022wam}, and in the $c$-jet case at LHCb such uncertainties may obscure discrimination among the FC models.
Pending better control of QCD uncertainties, we do not include the presently available $Z+c$ experimental data into our analysis.

In the meantime, available deep-inelastic scattering measurements have limited sensitivity to IC. While the EMC data~\cite{EuropeanMuon:1982xfn} on the charm structure function, $F^c_2$, have been suggested \cite{Hoffmann:1983ah,Harris:1995jx} as hinting at the
existence of IC, Refs.~\cite{Jimenez-Delgado:2014zga,Hou:2017khm} ran into difficulties fitting these data under a variety of physics scenarios. 
The CT14 IC analysis \cite{Hou:2017khm} examined the EMC data set in depth and did not include it in the published fit because these data did not follow modern standards in characterizing experimental systematics and were analyzed at LO. This study~\cite{Hou:2017khm} found $\chi^2/N_\mathrm{pt}\! \approx\! 2.3-3.5$ for the EMC data regardless of the IC model (cf.~$\chi^2/N_\mathrm{pt}\! =\! 4.3$ in Ref.~\cite{Jimenez-Delgado:2014zga}). In the current CT18 framework, we continue to find comparably high $\chi^2$ for the EMC $F_2^c$, as well as strong dependence on the prescription for the systematic uncertainty, which is not under adequate control for this data set.
The CT18 theory predictions {\it overshoot} the EMC data points over the whole $x$ range, likely reflecting the magnitude of the large-$x$ gluon PDF, except for the two highest-$Q$ points, for which inclusion of the FC leads to a few-unit improvement in $\chi^2$. In turn, the values of the gluon PDF at these $x$ and $Q$ reflect constraints primarily from well-fitted LHC and Tevatron jet experiments. Hence, the CT18 FC PDFs do not include the EMC $F_2^c$ data. 

SLAC DIS measurements on both the proton and
deuteron imposed stringent limits on any allowed FC in Ref.~\cite{Jimenez-Delgado:2014zga}, which may be modified nevertheless by considering a wider range of PDF and higher-twist parametrizations. (Note also that IC is nominally an NNLO effect in $\alpha_s$~\cite{Hou:2017khm}.)
In contrast, Ref.~\cite{Ball:2022qks} discussed in Sec.~5 obtained $\chi^2/N_\mathrm{pt}\! \gtrsim\! 1$ for the EMC data. Unlike CT, Ref.~\cite{Ball:2022qks} modified the nominal data reported by EMC. These include
a larger overall correlated systematic uncertainty (15\%, vs.~11\% as given in the original EMC publication); a shift in the central values by a factor of 0.82 based on a reduced branching ratio; and additional nuclear
uncertainties. These adjustments reduce the $\chi^2/N_\mathrm{pt}$ for NNPDF, as does a different behavior of the high-$x$ gluon relative to CT, which lowers the theoretical predictions over the whole range independently of the inclusion of FC.
The CT default prescription is to fit the EMC data set based on the reported experimental uncertainties; an in-depth look into the CT18 and NNPDF fits may shed light on the implications of the EMC $F^c_2$ data for FC.


\paragraph{\bf 4.~CT18 Fitted Charm PDFs}
\hypertarget{sec4}{The CT18 FC} family of PDFs presented in this article is obtained by repeating the CT18 NNLO and CT18X NNLO analyses using three parametrizations of the charm and anticharm PDFs at the initial scale, $Q_0=1.265\mbox{ GeV}$ (slightly below the charm pole mass, $m_c=1.27\mbox{ GeV}$, where the transition from three to four active quark flavors in these fits takes place; we note that this was $m_c\! =\! 1.3$ GeV for the CT18 and CT18X NNLO fits). The CT18 FC analysis indicates that the constraining power of available data remains insufficient for establishing the shape and non-zero normalization of the FC PDFs derived from high-$x$ IC models. This conclusion is consistent
with the findings of the earlier CT14 IC study as well as Ref.~\cite{Jimenez-Delgado:2014zga}. 

\begin{figure}[tb]
\center
\includegraphics[width=0.33\textwidth]{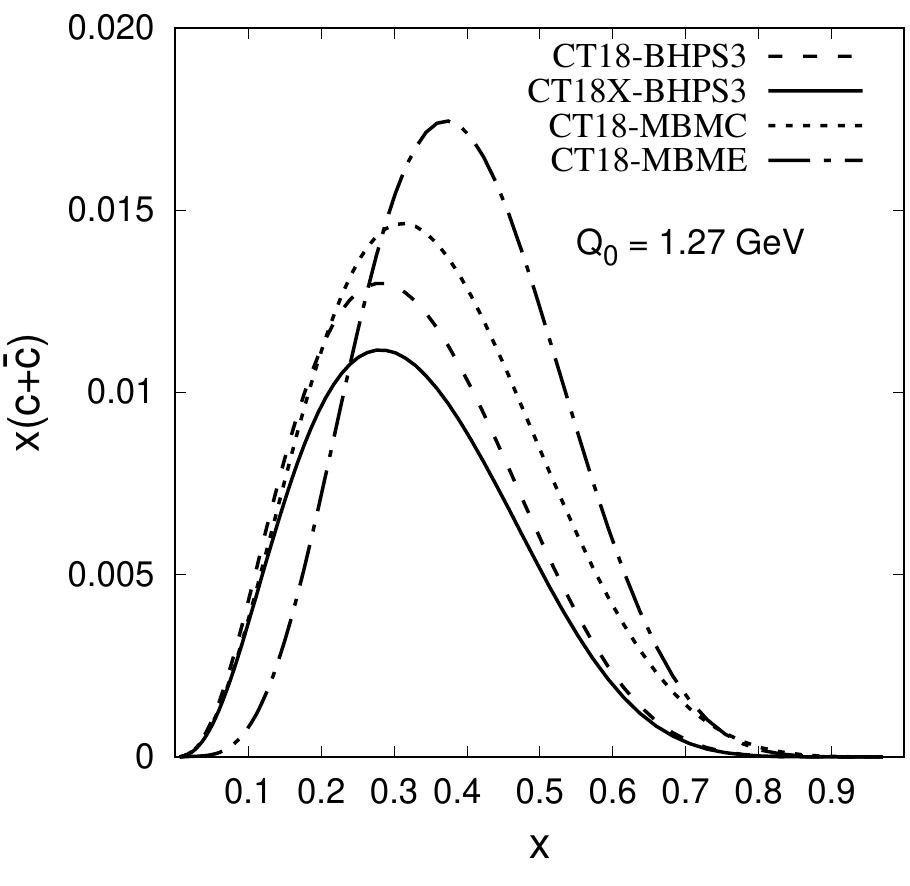}
\includegraphics[width=0.33\textwidth]{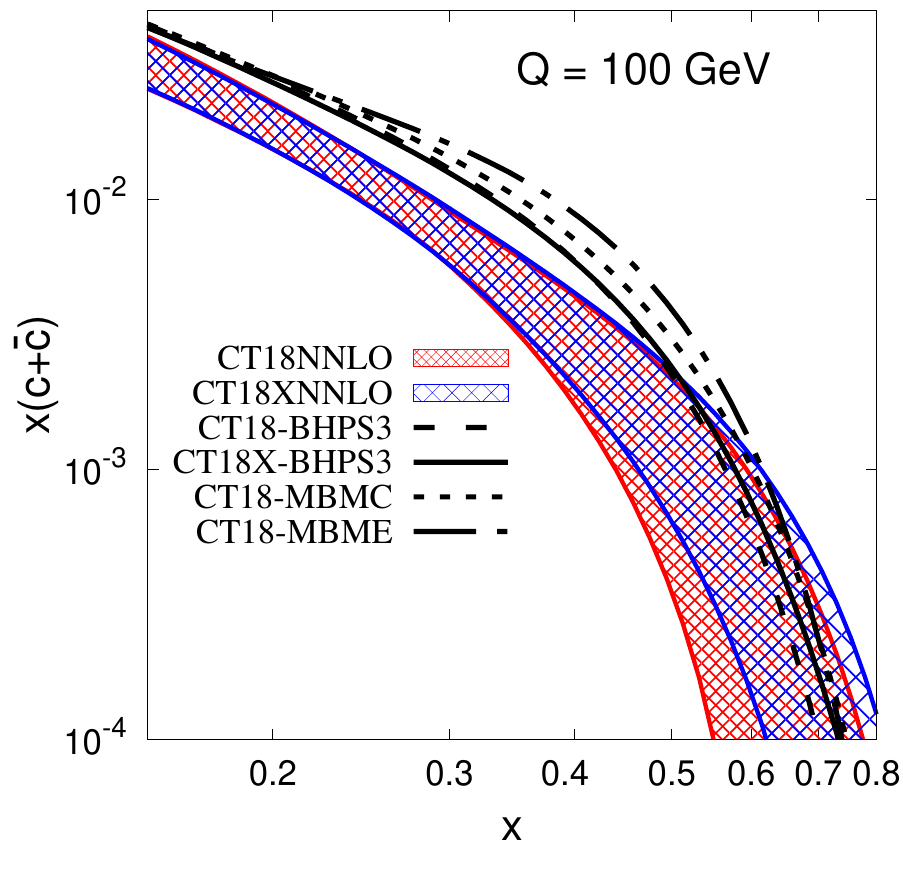}
\includegraphics[width=0.325\textwidth]{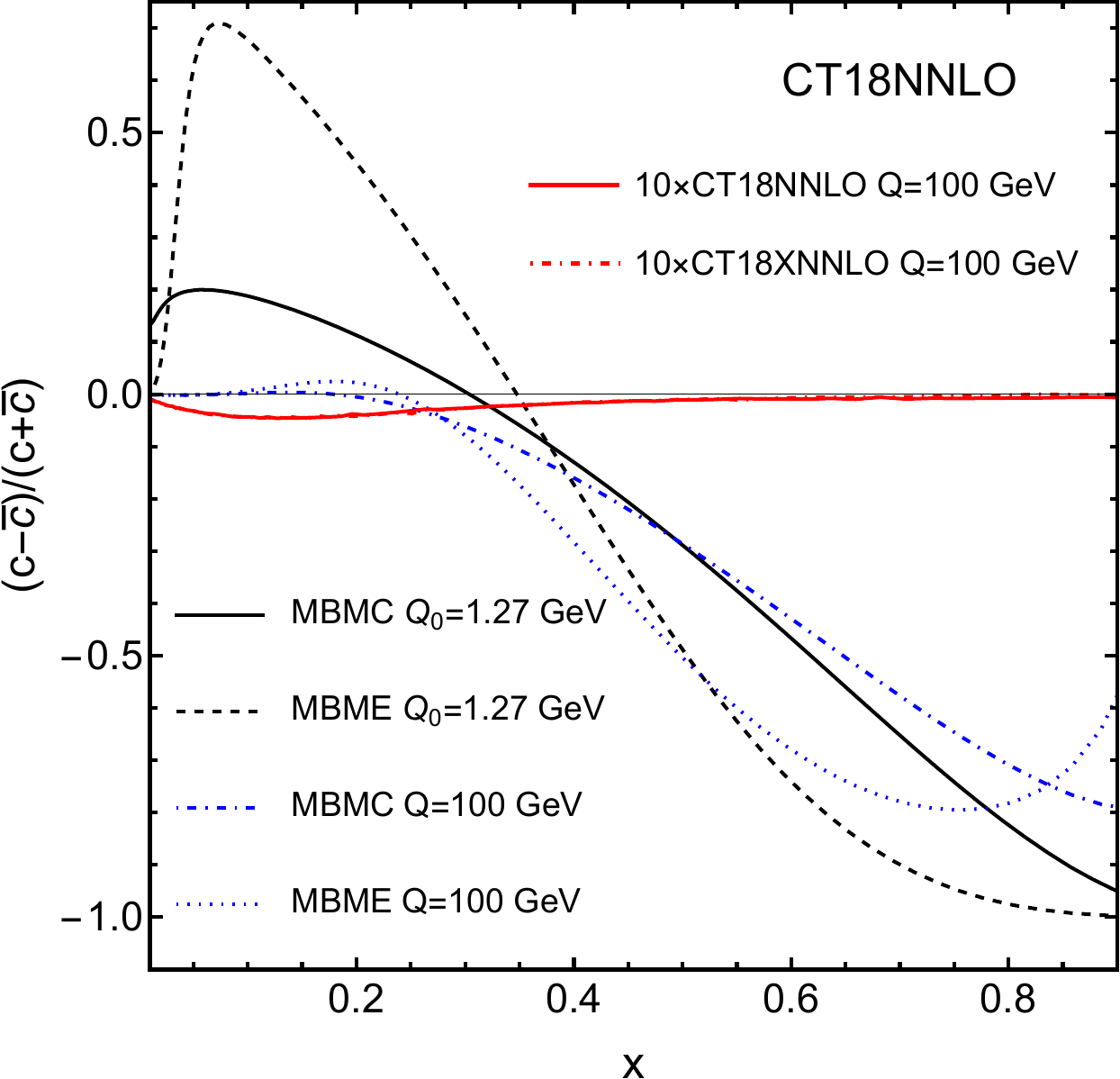}
	\caption{A comparison among the four FC models used in this study, CT18 and CT18X NNLO BHPS3 as well as CT18 NNLO MBMC/MBME. We show the fitted PDFs based on these for $xc^+(x,Q)$ at
	$Q\!=\!Q_0\!=\!1.27$ GeV (left) and $Q\!=\!100$ GeV (center). We also plot the asymmetric ratio, $c^-/c^+$, for MBMC and MBME scenarios at both	scales, as well as for CT18(X) extrinsic scenarios at $Q=100$ GeV augmented by a factor of ten (right).
\label{fig:cPDFs}}
\end{figure}

The CT18 FC study is based on the up-to-date NNLO QCD theory and data selections of the CT18~\cite{Hou:2019efy} global analysis as well as dedicated follow-up studies, including  CT18As\_Lat, which explored an $s\neq \bar s$ initial parametrization with the inclusion of lattice data~\cite{Hou:2022sdf}.
Collectively, the published CT18 FC ensemble encompasses 12 PDF sets: 4 variations of the analysis and underlying IC model --- specifically, the BHPS3 \cite{Blumlein:2015qcn,Hou:2017khm} and meson-baryon model (MBM)~\cite{Hobbs:2013bia} --- for each of which
we release 3 PDF sets corresponding to the central (best) fit and fits at intervals of $\Delta \chi^2\! =\! 10$ and $30$. The latter $\Delta \chi^2$
value approximates the standard $68\%$ C.L.~CT tolerance, while the former represents a more restrictive scenario compatible with the MSHT20 tolerance \cite{Bailey:2020ooq}. These PDFs thus estimate uncertainties in FC at high $x$ in accord with the common tolerance criteria.

As discussed in \hyperlink{sec2}{Sec.~2}, nonperturbative charm can be 
envisioned as an effective 5-quark Fock state \cite{Brodsky:1980pb,Pumplin:2005yf,Chang:2011vx,Hobbs:2013bia}, with the resulting charm-anticharm configuration largely
co-moving with the proton. The resulting IC PDF thus possesses a valence-like shape at $Q_0\! \sim\! m_c$, with an enhancement in $c(x,Q)$ at $x > 0.1$ which, for sufficiently large normalizations, survives above the perturbatively-generated charm PDF to electroweak scales, $Q^2 \gg m_c^2$ --- see the illustration in Fig.~4 of Ref.~\cite{Hou:2017khm}, as well as Fig.~\ref{fig:cPDFs} here.  The BHPS3 fits assume the BHPS3 IC model discussed in Ref.~\cite{Hou:2017khm} under two variants, namely, using the CT18 NNLO baseline and the alternative CT18X NNLO~\cite{Hou:2019efy} scenario in which DIS data are fitted using a Bjorken $x$-dependent factorization scale to model small-$x$ saturation. For the MBM, we consider two realizations based on confining (MBMC) and effective-mass (MBME) quark models as discussed in detail in Ref.~\cite{Hobbs:2013bia}. The two MBM models predict more significant differences in $c$ and $\bar c$ PDFs at $Q_0$, in contrast to the other scenarios in which $c\! \neq\! \bar{c}$ is generated only by higher-order perturbative corrections. 

Fig.~\ref{fig:cPDFs} plots the resulting $xc^+(x,Q)$ at $Q\!=\!1.27$ GeV ($\sim\! Q_0$) and $Q\!=\!100$ GeV in the left and center panels, as well as the asymmetric ratio, $c^-(x,Q)/c^+(x,Q)$, in the rightmost panel. These plots show that the considered models 
traverse a range of FC shapes at the initial scale, $Q_0$, with some differences surviving to the scales probed by high-energy experiments.

\begin{figure}[htb!]
\center
\includegraphics[width=0.6\textwidth]{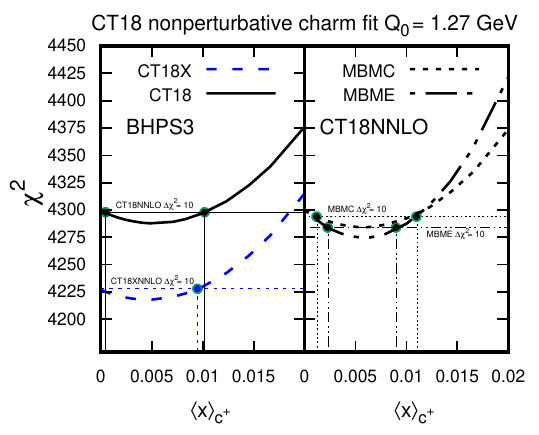}
\includegraphics[width=0.38\textwidth]{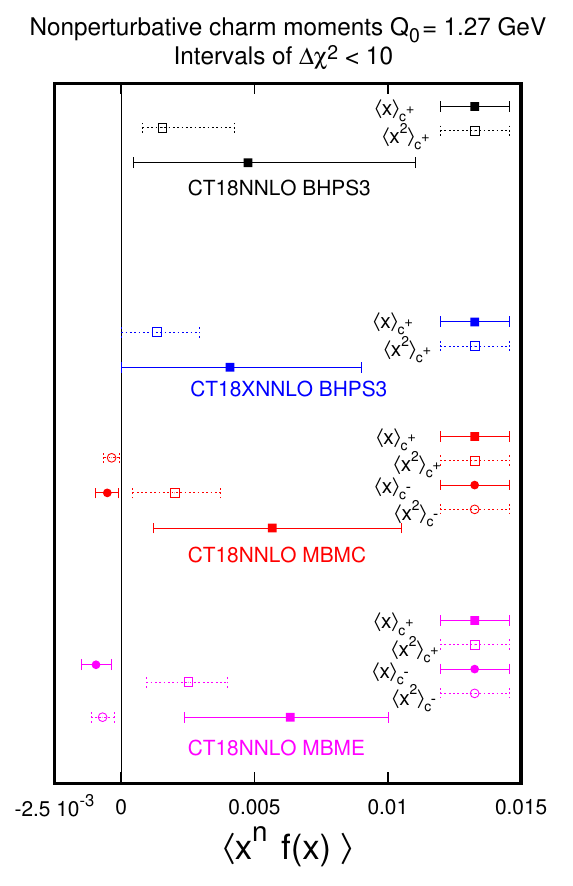}
\caption{(Left) Absolute $\chi^2$ values as a function of the charm momentum fraction, $\langle x \rangle_\mathrm{c^+}$, at the input scale, $Q = Q_0 = 1.27$ GeV, for each of the FC scenarios considered in this work: the BHPS3 model \cite{Blumlein:2015qcn,Hou:2017khm} within CT18 NNLO and CT18X NNLO and the
meson-baryon models (MBMs) implemented in CT18 NNLO which allow $c\! \neq\! \bar{c}$. (Right) The associated values of the available first and second moments
of the $c^\pm\! =\! c\! \pm\! \bar{c}$ PDF combinations for the normalizations preferred by each
fit are shown on the left. The intervals correspond to the increase in $\Delta\chi^2 < 10$ from the respective best-fit values. Of these moments, we note that $\langle x \rangle_\mathrm{c^+}$ and $\langle x^2 \rangle_\mathrm{c^-}$ are calculable in lattice
QCD.
\label{fig:FC-chi2}}
\end{figure}

Critically, these are the central fitted PDFs: the uncertainty of the
full normalization remains considerable, as found earlier~\cite{Hou:2017khm}.
Fig.~\ref{fig:FC-chi2} (left) quantifies a part of 
this uncertainty by the $\chi^2$ dependence
on $\langle x \rangle_{c^+}$. A very mild preference for 
$\langle x \rangle_\text{FC} \equiv \langle x \rangle_{c^+}(Q_0)\! \sim \! 0.5\%$ is seen in
all considered CT18 FC fits, marked by a $\Delta \chi^2\! \approx\! 10$-unit difference with respect to the
``no FC'' scenario. There is some disagreement among the fitted data sets regarding the FC magnitude, with the BCDMS and combined HERA DIS data exerting an upward pull on the total charm fraction, while the E866 $pp$ cross sections and 7, 8 TeV LHCb $W/Z$ data prefer
small FC. Fig.~\ref{fig:FC-chi2-LM} of the SD section dissects these pulls further via Lagrange Multiplier (LM) scans over $\langle x \rangle_\text{FC}$. The HERA charm-tagged DIS cross sections, $\sigma^c_r$, while influential in constraining the low-$x$ charm and gluon,
have apparently weaker pulls with respect to high-$x$ FC scenarios, and therefore do not appear in the plots of that figure.
{\bf Charm-anticharm asymmetry.} In addition to predicting a somewhat different high-$x$ behavior for the FC PDF relative to BHPS, the MBM involves hadronic interactions that break the $c\!=\!\bar{c}$ assumption at $Q_0$, unlike the BHPS model.
As an illustration, we demonstrate this using the MBM as a specific model case. We stress that the detailed $x$ dependence we obtain for $c\!-\bar{c}$ should not be taken as a strict prediction, but rather as an example of the asymmetric charm-anticharm scenarios to which future precise data might be sensitive. As discussed in Sec.~III of Ref.~\cite{Hobbs:2013bia}, MBMs typically involve the appearance of nonperturbative (anti)charm in virtual (meson) baryon states into which the proton is allowed to dissociate; the anticharm quark naturally carries a greater share of its parent meson's momentum relative to charm in the corresponding intermediate baryon, producing a harder high-$x$ distribution for $\bar{c}(x)$ relative to $c(x)$. When normalized to $c+\bar{c}$, the high-$x$ $c\!-\!\bar{c}$ asymmetry shown in Fig.~\ref{fig:cPDFs} (right) consequently approaches
$-1$ at $Q\!=\!Q_0$ and remains negative after evolving to higher energy scales.
Definite observation of a significant charm-anticharm asymmetry in either phenomenology or lattice calculations would be a substantial confirmation of the presence of IC.
If nonzero, considerably higher precision in both data and theory will be necessary to determine the sign, shape, and magnitude of $xc^-(x)$ in future QCD fits, much as we argue for the size and shape of $xc^+(x)$. 

The magnitudes of the FC and charm-anticharm asymmetry can be alternatively quantified by the symmetric ($c^+$) and asymmetric ($c^-$) first and second Mellin moments $\langle x^{1,2} \rangle$, which
we plot in Fig.~\ref{fig:FC-chi2} (right) as central values and uncertainties corresponding to the $\Delta \chi^2\! \leq\! 10$ intervals with respect to the $\chi^2$ minima in the left panel. The allowed ranges for each moment are quite substantial, being practically consistent with zero even based on the more restrictive $\Delta \chi^2\! =\! 10$ criterion. As for the approximate CT tolerance of $\Delta \chi^2=30$, we obtain $\langle x \rangle_\text{FC} \lesssim 0.013$ in all scenarios, with the precise uncertainties quoted in Eq.~(\ref{FC-moments}) of the SD section. This represents a moderate reduction in the allowed upper value of $\langle x \rangle_\text{FC} \lesssim 0.02$ obtained based on the BHPS3 model fitted in the CT14 IC study~\cite{Hou:2017khm}.
The negative $c^-$ of the MBM is reflected in the negative values we obtain for the asymmetric moments, $\langle x^{1,2} \rangle_{c^-}$, as also shown in Fig.~\ref{fig:FC-chi2}. We point out that only some moments in Fig.~\ref{fig:FC-chi2} can be computed by lattice QCD techniques: specifically, $\langle x \rangle_{c^+}$ and $\langle x^2 \rangle_{c^-}$. Although lattice QCD calculations remain
at an early stage for quantities related to charm, precise information on either of these moments, or complementary calculations of the high-$x$
charm PDF, would be very useful for constraining possible FC scenarios.

One might reason that the nonperturbative symmetry-breaking mechanism(s) that produce $c\!-\!\bar{c} \neq 0$ should have some analogue in the strange sector; we have considered such a possibility by
performing alternative fits of MBMC(E) starting from a variant fit (CT18As2) from Ref.~\cite{Hou:2022sdf} as a baseline, which allowed $s\!\neq\!\bar{s}$. While we see indications of mild correlations between the strange and charm PDFs in these fits, with the inclusion
of FC according to MBMC(E) causing small reductions in the high-$x$ strange (and gluon) PDF, $s\!-\!\bar{s}$ remains largely unaffected; hence, even in fits with a strange-antistrange asymmetry, we
obtain similar results for the FC PDFs themselves.
We include several plots illustrating the PDFs obtained in these simultaneous fits with $s\!\neq\!\bar{s}$ and $c\!\neq\!\bar{c}$ in the SD.

The improvement of $\chi^2$ by no more than 25 units 
for $\langle x\rangle_\text{FC}\approx 0.5\%$ in all considered FC scenarios in Fig.~\ref{fig:FC-chi2} is milder than in the CT14 IC analysis, where $\langle x\rangle_\text{FC}\approx 0.8-1\%$ corresponded to a reduction in $\chi^2$ of up to $\approx \text{40}$ units for the BHPS models, cf.~Fig.~5 in Ref.~\cite{Hou:2017khm}. 
Ultimately, the very shallow preferences for FC in Fig.~\ref{fig:FC-chi2}  comply with the findings of other PDF fitting efforts, including the observation by MSHT that, when PDFs are fitted at partial N$^3$LO$'$~\cite{McGowan:2022nag}, there can be an enhancement in the perturbatively-generated charm PDF at high $x$, a feature which reduces the parameter space available for nonperturbative charm. We also note that the constraints we obtain on large-$x$ FC, while already shallow, depend only mildly on $m_c^\mathit{pole}$ --- the parameter whose best-fit value compensates in part for the missing N$^3$LO charm-quark scattering contribution to DIS cross sections \cite{Gao:2013wwa} --- and the form of the gluon parametrization~\cite{Hou:2017khm}.
When coupled with formal ambiguities in the relation of IC to FC PDFs, the fairly weak and incoherent pulls of the few experiments suggest a clear need for better experimental constraints.

\paragraph{\bf 5.~NNPDF 2022 IC analysis}
\hypertarget{sec5}{In contrast,} the NNPDF group has recently claimed~\cite{Ball:2022qks} robust, $3\sigma$ evidence for ``intrinsic charm" --- technically, FC by the definition above --- using an $x$-dependent deviation of their FC PDF from a ``no-FC'' scenario, up to a PDF uncertainty assessed using the default NNPDF framework. That this local criterion can serve as a robust hypothesis test for the presence of FC is not necessarily obvious, as it may be subject to fluctuations in the fitted PDFs. The statistical significance of NNPDF's quantitative findings depends both on the magnitude of their central FC PDF at the comparison scale, $Q = m_c=1.51 $ GeV, and on the probability levels assigned to the PDF uncertainties. NNPDF estimates both based on an ensemble of trained Monte-Carlo replicas, rendering a mean (central) NNPDF FC PDF that is enhanced at $x\!>\!0.2$ in a bump-like pattern. With this, the new NNPDF analysis confirms the earlier observation of CT10 and CT14 IC studies (see especially Figs.~4 and~14 in \cite{Hou:2017khm}) that a BHPS-like FC with a momentum fraction of $0.5\!-\!1\%$ describes the observations as well as the purely perturbative charm, if not marginally better.  

The crucial distinction with the CT18 FC analysis is that NNPDF's substantially narrower nominal PDF uncertainties seem to disfavor a ``no-FC" PDF at $x>0.2$. A recent publication \cite{Courtoy:2022ocu} critically assessed these uncertainties 
using the publicly-released NNPDF4.0 fitting code to conclude that the NNPDF4.0 uncertainties on FC are likely underestimated. Specifically, figures in Sec.~3.E of Ref.~\cite{Courtoy:2022ocu} make evident that the NNPDF4.0 analysis would actually allow solutions with (nearly) zero FC with high probability under a more comprehensive sampling. The publication discusses the reasons why the effective prior introduced by the NNPDF replica training may spuriously omit these well-behaved and therefore acceptable solutions, following the usual practice in CT PDF analyses. Furthermore, Ref.~\cite{Courtoy:2022ocu} finds that the large-$x$ FC and $s\!-\!\bar{s}$ PDFs are correlated, introducing an additional ambiguity in the region of $x\sim 0.4$ where separation from rapidly falling $\bar u$, $\bar d$, $s$, and $\bar s$ PDFs is particularly challenging. When the extra small-FC solutions are included, the MC PDF uncertainty alone washes out the evidence for a non-zero FC at large $x$. 
This uncertainty is further increased by accounting for the possible differences in approximating the experimental systematic uncertainties.

Another statistical indicator of FC, based on the charm PDF's first moment, $\langle x\rangle_\text{FC}$, at $Q\! =\! m_c\! =\! 1.51$ GeV, is affected by a large missing higher-order uncertainty (MHOU) evident at $x<0.4$ in Fig.~1 of \cite{Ball:2022qks}. At $x<0.1$, the central NNPDF4.0 PDF predicts a large and negative FC that is difficult to reconcile with the valence-like shape of nonperturbative IC models. The negative low-$x$ behavior of the central FC spotlights the tenuous connection of NNPDF's FC solutions to the nonperturbative models that do not favor this. On the other hand, the MHOU remains large at $x<0.1$: including the N$^3$LO matching coefficients in the PDF evolution, without consistently including N$^3$LO terms in the coefficient functions, does not genuinely reduce the scale uncertainty. Ref.~\cite{Ball:2022qks} determines the first moment to be $\langle x \rangle_\mathrm{FC} = 0.62 \pm 0.28\%$ based on the PDF uncertainty (PDFU) only and $0.62 \pm 0.61\%$ after adding the MHOU. The latter, in essence, makes $\langle x \rangle_\mathrm{FC}$ consistent with zero at $1\sigma$.

For the interpretation of these results, it is worth noting that CT18 and NNPDF4.0 employ very similar NNLO theoretical frameworks. Not only CT and NNPDF use respectively the SACOT-$\chi$ and FONLL-C factorization schemes that are perturbatively equivalent up to NNLO, these schemes also produce numerically close predictions for key cross sections, as has been demonstrated repeatedly in joint benchmarking exercises such as the recent one in Ref.~\cite{PDF4LHCWorkingGroup:2022cjn}.
As one of the schemes of the ACOT family~\cite{Aivazis:1993pi}, the SACOT-$\chi$ scheme achieves the same level of accuracy as other ACOT approaches, while using simpler approximations. The derivation of the SACOT-$\chi$ scheme to NNLO~\cite{Guzzi:2011ew} and its application to the FC scenario~\cite{Hou:2017khm} demonstrate that, when including a power-suppressed FC PDF of the kind encountered for the nucleon, predictions of the original (``full'') ACOT and SACOT-$\chi$ schemes agree up to terms of order $\Lambda^2/Q^2$, {\it i.e.}, to an accuracy exceeding the validity of the factorization theorem. Therefore, the SACOT-$\chi$ and FONLL-C schemes, being perturbatively equivalent to the full ACOT scheme, are of the same accuracy.
On the other hand, it has been established that the NNPDF3.1 and, even more so, NNPDF4.0 methodologies may produce smaller PDF error bands compared to either CT18 or MSHT20, even when fitting to a similar data set \cite{PDF4LHCWorkingGroup:2022cjn}; in this case, the larger PDF error estimates of the CT or MSHT groups may decrease the statistical significance of the observed signal.

Regarding the setup relevant to the FC study, we point out that the CT14/CT18 and NNPDF4.0 procedures for introducing FC and evolving it over mass thresholds are equivalent, with the only distinction being that CT parametrizes FC at a scale $Q_0$ slightly below $m_c$ in the $N_f\!=\!3$ scheme and evolves it to higher $Q$ by matching to the $N_f\!=\!4$ scheme at $Q\!=\!m_c$. The NNPDF4.0 analysis, on the other hand, parametrizes FC at $Q_0 (=1.65\,\mathrm{GeV})$ above $m_c$ in the $N_f\!=\!4$ scheme and evolves it backwards to $Q\!=\!m_c\!=\!1.51$ GeV, where the resulting FC is converted into the $N_f\!=\!3$ scheme and presented. Thus, the CT18 and NNPDF4.0 FC parametrizations at a scale $Q \leq m_c$ can be directly compared at NNLO. 

By admitting the enlarged PDFU+MHOU uncertainty of the NNPDF4.0 analysis, we arrive at a general consensus between the latest CT18 and NNPDF findings with regard to FC. Namely, FC may possibly improve $\chi^2$ or a related figure of merit, although with low confidence that does not rise to the evidence level. In the same vein, the observed behavior of the NNPDF's FC would be easier to reconcile with nonperturbative IC models once these larger uncertainties are considered. We already pointed out that the negative central FC at $x<0.1$ and valence-like FC shape of the models can be reconciled by accounting for a large higher-order uncertainty. On the other hand, at $x>0.1$, the central NNPDF FC PDF, $xc^+(x, Q\!=\!1.51\,\mathrm{GeV})$, peaks at $x\!\gtrsim\! 0.4$; this is considerably harder than the shapes that many nonperturbative models naturally produce: {\it e.g.}, the MBMs of Ref.~\cite{Hobbs:2013bia} generally peak in the region of $0.3\!\lesssim \! x\! \lesssim 0.4$ or below, as does the BHPS model~\cite{Brodsky:1980pb}, assuming conventional, $\mathcal{O}(100\,\mathrm{MeV}$) constituent quark masses. (We point out that the slightly higher scale, $Q\! =\! 1.51\! >\! 1.27$ GeV, at which NNPDF compute their FC PDFs, make the hard shape even more striking.) 
While the neural network approach notably entails a highly flexible parametrization beyond the individual model-based forms in CT18FC, we have explored
a range of high-$x$ shapes and peak locations for FC as shown in Fig.~\ref{fig:cPDFs} (left), with no indication of significant $\chi^2$ dependence associated
with these variations in Fig.~\ref{fig:FC-chi2} (left). These findings echo the conclusion in Ref.~\cite{Hou:2017khm}, wherein the left panel of Fig.~6 illustrated the very 
weak dependence of $\chi^2$ on the charm mass, which controls the position of the peak of FC for BHPS-like models.
Moreover, although IC models might be fine-tuned to produce harder shapes beyond those obtained with natural parameter choices, such nominal differences again highlight the formal theory development needed to relate FC to IC as argued in~\hyperlink{sec2}{Sec.~2}. These discrepancies are also relieved by assuming larger uncertainties. The only guaranteed way to improve the perturbative uncertainty is to fully implement radiative contributions of the same order in $\alpha_s$. For DIS cross sections, the current uncertainties can be reduced by fully implementing the N$^3$LO contributions, if massive-quark terms are included. 
For the LHCb $(Z+c)/(Z+\mbox{jets})$ ratio, the currently formidable uncertainty in predictions that we examine in the SD must be better controlled through a combination of computations beyond NLO+PS and advances in flavored jet algorithms.


%
\paragraph{\bf 6.~Conclusions}
\hypertarget{sec6}{The new CT18 NNLO FC global analysis} concludes that evidence for nonperturbative charm continues to be elusive, counter to the recent finding in Ref.~\cite{Ball:2022qks}. As such, the subject remains open and in need of further theory and experimental data.
Specifically, a number of complementary developments are required if future PDF analyses are to be capable of discriminating nonzero IC
with high statistical confidence:
(1) theoretical progress relating nonperturbative correlation functions in factorization-based calculations to wave functions
formulated in terms of nonperturbative charm; this includes separating contributions from twist-4 and beyond as discussed in Ref.~\cite{Hou:2017khm};
(2) clean and sensitive experimental data at hadron-hadron and lepton-hadron colliders to test the ``universality" of the IC PDF;
(3) theoretical calculations at NNLO, and possibly including parton-showering effects, for relevant data like $Z+c$ production at the LHC, to correctly extract both the central value and uncertainty of the FC PDFs; and
(4) faithful estimates of PDF errors in global analyses. (See Refs.~\cite{PDF4LHCWorkingGroup:2022cjn,Courtoy:2022ocu}.) This more comprehensive
uncertainty quantification entails improved understanding of PDF correlations.
This will be indispensable, as the size of IC PDF at high $x$ is strongly correlated with the high-$x$ gluon due to the momentum sum rule.
Since NNPDF's high-$x$ gluon is smaller than that of CT and MSHT, NNPDF is more likely to obtain a comparative enhancement in their FC PDF at larger values of $x$. Joint PDF benchmarking exercises like those in Ref.~\cite{PDF4LHCWorkingGroup:2022cjn} could include FC parametrizations to understand these issues. 

Regarding Point (2), direct EIC data similar to the EMC measurements of the charm structure function, $F^c_2$, at high $x$ and over a range of $Q^2$
values, would be invaluable for revisiting the possible large-$x$ excess suggested by CT14 BHPS, NNPDF, and the traditional nonperturbative charm
models~\cite{GuzziIC:2011,Hobbs:2017fom}. Moreover, manifestations of IC might affect the physics program at the proposed CERN Forward Physics Facility~\cite{Anchordoqui:2021ghd,Feng:2022inv}. Knowledge of the $Q^2$ dependence of this quantity would allow tests to unravel the nature of nonperturbative charm against
power-suppressed contributions arising in a twist expansion.
Additional discriminating inputs include possible lattice QCD calculations of $\langle x \rangle_{c^+}$ and/or $\langle x^2 \rangle_{c^-}$; these
would be highly informative from the perspective of having an independent determination of the total charm magnitude in the first case. In the second
case, a measure of the possible $c$, $\bar{c}$ asymmetry would be of great interest given the fact that this asymmetry, if non-negligible, would
principally originate from nonperturbative dynamics~\cite{Hobbs:2013bia,Sufian:2020coz}. These inputs might be augmented by $x$-dependent lattice information from
the quasi- or pseudo-PDF methods~\cite{Ji:2013dva,Radyushkin:2017cyf,Zhang:2020dkn}.

We will make available 12 grids for the CT18 FC NNLO PDFs described above as a part of the LHAPDF library (\url{https://lhapdf.hepforge.org/}) and at the CTEQ-TEA website (\url{https://ct.hepforge.org/}).

\vspace{1\baselineskip}{\bf Acknowledgments.}
We thank Aurore Courtoy, Yao Fu, Tie-Jiun Hou, and other CTEQ-TEA members for discussions.
The work of MG is supported by the National Science Foundation under Grant No.~PHY-2112025.
The work of TJH at Argonne National Laboratory was supported by the U.S.~Department of Energy, Office of Science, under Contract No.~DE-AC02-06CH11357.
PMN is partially supported by the U.S. Department of Energy under Grant No.~DE-SC0010129 and by the Fermilab URA award, using the resources of the Fermi National Accelerator Laboratory (Fermilab), a U.S. Department of Energy, Office of Science, HEP User Facility. Fermilab is managed by Fermi Research Alliance, LLC (FRA), acting under Contract No. DE-AC02-07CH11359.
The work of KX is supported by the U.S. Department
of Energy under grant No. DE-SC0007914, the U.S. National Science Foundation under Grants No. PHY-1820760 and also in part by PITT PACC.
CPY was supported by the U.S. National Science Foundation under Grant No.~PHY-2013791 as well as the Wu-Ki Tung endowed chair in particle physics.

\bibliographystyle{elsarticle-num-names}
\bibliography{ct18bibtex}

\begin{thebibliography}{45}
\expandafter\ifx\csname natexlab\endcsname\relax\def\natexlab#1{#1}\fi
\providecommand{\url}[1]{\texttt{#1}}
\providecommand{\href}[2]{#2}
\providecommand{\path}[1]{#1}
\providecommand{\DOIprefix}{doi:}
\providecommand{\ArXivprefix}{arXiv:}
\providecommand{\URLprefix}{URL: }
\providecommand{\Pubmedprefix}{pmid:}
\providecommand{\doi}[1]{\href{http://dx.doi.org/#1}{\path{#1}}}
\providecommand{\Pubmed}[1]{\href{pmid:#1}{\path{#1}}}
\providecommand{\bibinfo}[2]{#2}
\ifx\xfnm\relax \def\xfnm[#1]{\unskip,\space#1}\fi
\bibitem[{Brodsky et~al.(1980)Brodsky, Hoyer, Peterson, and
  Sakai}]{Brodsky:1980pb}
\bibinfo{author}{S.~J. Brodsky}, \bibinfo{author}{P.~Hoyer},
  \bibinfo{author}{C.~Peterson}, \bibinfo{author}{N.~Sakai},
\newblock \bibinfo{title}{{The Intrinsic Charm of the Proton}},
\newblock \bibinfo{journal}{Phys. Lett.} \bibinfo{volume}{B93}
  (\bibinfo{year}{1980}) \bibinfo{pages}{451--455}.
  \DOIprefix\doi{10.1016/0370-2693(80)90364-0}.
\bibitem[{Pumplin et~al.(2007)Pumplin, Lai, and Tung}]{Pumplin:2007wg}
\bibinfo{author}{J.~Pumplin}, \bibinfo{author}{H.-L. Lai},
  \bibinfo{author}{W.-K. Tung},
\newblock \bibinfo{title}{{The Charm Parton Content of the Nucleon}},
\newblock \bibinfo{journal}{Phys. Rev.} \bibinfo{volume}{D75}
  (\bibinfo{year}{2007}) \bibinfo{pages}{054029}.
  \DOIprefix\doi{10.1103/PhysRevD.75.054029}.
  \href{http://arxiv.org/abs/hep-ph/0701220}{{\tt arXiv:hep-ph/0701220}}.
\bibitem[{Jimenez-Delgado et~al.(2015)Jimenez-Delgado, Hobbs, Londergan, and
  Melnitchouk}]{Jimenez-Delgado:2014zga}
\bibinfo{author}{P.~Jimenez-Delgado}, \bibinfo{author}{T.~J. Hobbs},
  \bibinfo{author}{J.~T. Londergan}, \bibinfo{author}{W.~Melnitchouk},
\newblock \bibinfo{title}{{New limits on intrinsic charm in the nucleon from
  global analysis of parton distributions}},
\newblock \bibinfo{journal}{Phys. Rev. Lett.} \bibinfo{volume}{114}
  (\bibinfo{year}{2015}) \bibinfo{pages}{082002}.
  \DOIprefix\doi{10.1103/PhysRevLett.114.082002}.
  \href{http://arxiv.org/abs/1408.1708}{{\tt arXiv:1408.1708}}.
\bibitem[{Hou et~al.(2018)Hou, Dulat, Gao, Guzzi, Huston, Nadolsky, Schmidt,
  Winter, Xie, and Yuan}]{Hou:2017khm}
\bibinfo{author}{T.-J. Hou}, \bibinfo{author}{S.~Dulat},
  \bibinfo{author}{J.~Gao}, \bibinfo{author}{M.~Guzzi},
  \bibinfo{author}{J.~Huston}, \bibinfo{author}{P.~Nadolsky},
  \bibinfo{author}{C.~Schmidt}, \bibinfo{author}{J.~Winter},
  \bibinfo{author}{K.~Xie}, \bibinfo{author}{C.~P. Yuan},
\newblock \bibinfo{title}{{CT14 Intrinsic Charm Parton Distribution Functions
  from CTEQ-TEA Global Analysis}},
\newblock \bibinfo{journal}{JHEP} \bibinfo{volume}{02} (\bibinfo{year}{2018})
  \bibinfo{pages}{059}. \DOIprefix\doi{10.1007/JHEP02(2018)059}.
  \href{http://arxiv.org/abs/1707.00657}{{\tt arXiv:1707.00657}}.
\bibitem[{Ball et~al.(2022)Ball, Candido, Cruz-Martinez, Forte, Giani, Hekhorn,
  Kudashkin, Magni, and Rojo}]{Ball:2022qks}
\bibinfo{author}{R.~D. Ball}, \bibinfo{author}{A.~Candido},
  \bibinfo{author}{J.~Cruz-Martinez}, \bibinfo{author}{S.~Forte},
  \bibinfo{author}{T.~Giani}, \bibinfo{author}{F.~Hekhorn},
  \bibinfo{author}{K.~Kudashkin}, \bibinfo{author}{G.~Magni},
  \bibinfo{author}{J.~Rojo} (\bibinfo{collaboration}{NNPDF}),
\newblock \bibinfo{title}{{Evidence for intrinsic charm quarks in the proton}},
\newblock \bibinfo{journal}{Nature} \bibinfo{volume}{608}
  (\bibinfo{year}{2022}) \bibinfo{pages}{483--487}.
  \DOIprefix\doi{10.1038/s41586-022-04998-2}.
  \href{http://arxiv.org/abs/2208.08372}{{\tt arXiv:2208.08372}}.
\bibitem[{Brodsky et~al.(1984)Brodsky, Collins, Ellis, Gunion, and
  Mueller}]{Brodsky:1984nx}
\bibinfo{author}{S.~J. Brodsky}, \bibinfo{author}{J.~C. Collins},
  \bibinfo{author}{S.~D. Ellis}, \bibinfo{author}{J.~F. Gunion},
  \bibinfo{author}{A.~H. Mueller},
\newblock \bibinfo{title}{{Intrinsic Chevrolets at the SSC}},
\newblock in: \bibinfo{booktitle}{{1984 DPF Summer Study on the Design and
  Utilization of the Superconducting Super Collider (SSC) (Snowmass 84)}},
  \bibinfo{year}{1984}, p. \bibinfo{pages}{227}.
\bibitem[{Chang and Peng(2011)}]{Chang:2011vx}
\bibinfo{author}{W.-C. Chang}, \bibinfo{author}{J.-C. Peng},
\newblock \bibinfo{title}{{Flavor Asymmetry of the Nucleon Sea and the
  Five-Quark Components of the Nucleons}},
\newblock \bibinfo{journal}{Phys. Rev. Lett.} \bibinfo{volume}{106}
  (\bibinfo{year}{2011}) \bibinfo{pages}{252002}.
  \DOIprefix\doi{10.1103/PhysRevLett.106.252002}.
  \href{http://arxiv.org/abs/1102.5631}{{\tt arXiv:1102.5631}}.
\bibitem[{Hobbs et~al.(2017)Hobbs, Alberg, and Miller}]{Hobbs:2017fom}
\bibinfo{author}{T.~J. Hobbs}, \bibinfo{author}{M.~Alberg},
  \bibinfo{author}{G.~A. Miller},
\newblock \bibinfo{title}{{Bayesian analysis of light-front models and the
  nucleon\textquoteright{}s charmed sigma term}},
\newblock \bibinfo{journal}{Phys. Rev. D} \bibinfo{volume}{96}
  (\bibinfo{year}{2017}) \bibinfo{pages}{074023}.
  \DOIprefix\doi{10.1103/PhysRevD.96.074023}.
  \href{http://arxiv.org/abs/1707.06711}{{\tt arXiv:1707.06711}}.
\bibitem[{Pumplin(2006)}]{Pumplin:2005yf}
\bibinfo{author}{J.~Pumplin},
\newblock \bibinfo{title}{{Light-cone models for intrinsic charm and bottom}},
\newblock \bibinfo{journal}{Phys. Rev. D} \bibinfo{volume}{73}
  (\bibinfo{year}{2006}) \bibinfo{pages}{114015}.
  \DOIprefix\doi{10.1103/PhysRevD.73.114015}.
  \href{http://arxiv.org/abs/hep-ph/0508184}{{\tt arXiv:hep-ph/0508184}}.
\bibitem[{Hobbs et~al.(2014)Hobbs, Londergan, and Melnitchouk}]{Hobbs:2013bia}
\bibinfo{author}{T.~J. Hobbs}, \bibinfo{author}{J.~T. Londergan},
  \bibinfo{author}{W.~Melnitchouk},
\newblock \bibinfo{title}{{Phenomenology of nonperturbative charm in the
  nucleon}},
\newblock \bibinfo{journal}{Phys. Rev.} \bibinfo{volume}{D89}
  (\bibinfo{year}{2014}) \bibinfo{pages}{074008}.
  \DOIprefix\doi{10.1103/PhysRevD.89.074008}.
  \href{http://arxiv.org/abs/1311.1578}{{\tt arXiv:1311.1578}}.
\bibitem[{Ball et~al.(2017)}]{NNPDF:2017mvq}
\bibinfo{author}{R.~D. Ball}, et~al. (\bibinfo{collaboration}{NNPDF}),
\newblock \bibinfo{title}{{Parton distributions from high-precision collider
  data}},
\newblock \bibinfo{journal}{Eur. Phys. J. C} \bibinfo{volume}{77}
  (\bibinfo{year}{2017}) \bibinfo{pages}{663}.
  \DOIprefix\doi{10.1140/epjc/s10052-017-5199-5}.
  \href{http://arxiv.org/abs/1706.00428}{{\tt arXiv:1706.00428}}.
\bibitem[{Ball et~al.(2016)Ball, Bertone, Bonvini, Carrazza, Forte, Guffanti,
  Hartland, Rojo, and Rottoli}]{Ball:2016neh}
\bibinfo{author}{R.~D. Ball}, \bibinfo{author}{V.~Bertone},
  \bibinfo{author}{M.~Bonvini}, \bibinfo{author}{S.~Carrazza},
  \bibinfo{author}{S.~Forte}, \bibinfo{author}{A.~Guffanti},
  \bibinfo{author}{N.~P. Hartland}, \bibinfo{author}{J.~Rojo},
  \bibinfo{author}{L.~Rottoli} (\bibinfo{collaboration}{NNPDF}),
\newblock \bibinfo{title}{{A Determination of the Charm Content of the
  Proton}},
\newblock \bibinfo{journal}{Eur. Phys. J.} \bibinfo{volume}{C76}
  (\bibinfo{year}{2016}) \bibinfo{pages}{647}.
  \DOIprefix\doi{10.1140/epjc/s10052-016-4469-y}.
  \href{http://arxiv.org/abs/1605.06515}{{\tt arXiv:1605.06515}}.
\bibitem[{Aubert et~al.(1983)}]{EuropeanMuon:1982xfn}
\bibinfo{author}{J.~J. Aubert}, et~al. (\bibinfo{collaboration}{European
  Muon}),
\newblock \bibinfo{title}{{Production of charmed particles in 250-GeV $\mu^+$ -
  iron interactions}},
\newblock \bibinfo{journal}{Nucl. Phys. B} \bibinfo{volume}{213}
  (\bibinfo{year}{1983}) \bibinfo{pages}{31--64}.
  \DOIprefix\doi{10.1016/0550-3213(83)90174-8}.
\bibitem[{Hoffmann and Moore(1983)}]{Hoffmann:1983ah}
\bibinfo{author}{E.~Hoffmann}, \bibinfo{author}{R.~Moore},
\newblock \bibinfo{title}{{Subleading Contributions to the Intrinsic Charm of
  the Nucleon}},
\newblock \bibinfo{journal}{Z. Phys. C} \bibinfo{volume}{20}
  (\bibinfo{year}{1983}) \bibinfo{pages}{71}.
  \DOIprefix\doi{10.1007/BF01577720}.
\bibitem[{Ball et~al.(2022)}]{PDF4LHCWorkingGroup:2022cjn}
\bibinfo{author}{R.~D. Ball}, et~al. (\bibinfo{collaboration}{PDF4LHC Working
  Group}),
\newblock \bibinfo{title}{{The PDF4LHC21 combination of global PDF fits for the
  LHC Run III}},
\newblock \bibinfo{journal}{J. Phys. G} \bibinfo{volume}{49}
  (\bibinfo{year}{2022}) \bibinfo{pages}{080501}.
  \DOIprefix\doi{10.1088/1361-6471/ac7216}.
  \href{http://arxiv.org/abs/2203.05506}{{\tt arXiv:2203.05506}}.
\bibitem[{Boettcher et~al.(2016)Boettcher, Ilten, and
  Williams}]{Boettcher:2015sqn}
\bibinfo{author}{T.~Boettcher}, \bibinfo{author}{P.~Ilten},
  \bibinfo{author}{M.~Williams},
\newblock \bibinfo{title}{{Direct probe of the intrinsic charm content of the
  proton}},
\newblock \bibinfo{journal}{Phys. Rev. D} \bibinfo{volume}{93}
  (\bibinfo{year}{2016}) \bibinfo{pages}{074008}.
  \DOIprefix\doi{10.1103/PhysRevD.93.074008}.
  \href{http://arxiv.org/abs/1512.06666}{{\tt arXiv:1512.06666}}.
\bibitem[{Bailas and Goncalves(2016)}]{Bailas:2015jlc}
\bibinfo{author}{G.~Bailas}, \bibinfo{author}{V.~P. Goncalves},
\newblock \bibinfo{title}{{Phenomenological implications of the intrinsic charm
  in the $Z$ boson production at the LHC}},
\newblock \bibinfo{journal}{Eur. Phys. J. C} \bibinfo{volume}{76}
  (\bibinfo{year}{2016}) \bibinfo{pages}{105}.
  \DOIprefix\doi{10.1140/epjc/s10052-016-3941-z}.
  \href{http://arxiv.org/abs/1512.06007}{{\tt arXiv:1512.06007}}.
\bibitem[{Gauld et~al.(2020)Gauld, Gehrmann-De~Ridder, Glover, Huss, and
  Majer}]{Gauld:2020deh}
\bibinfo{author}{R.~Gauld}, \bibinfo{author}{A.~Gehrmann-De~Ridder},
  \bibinfo{author}{E.~W.~N. Glover}, \bibinfo{author}{A.~Huss},
  \bibinfo{author}{I.~Majer},
\newblock \bibinfo{title}{{Predictions for $Z$ -Boson Production in Association
  with a $b$-Jet at $\mathcal {O}(\alpha_s^3)$}},
\newblock \bibinfo{journal}{Phys. Rev. Lett.} \bibinfo{volume}{125}
  (\bibinfo{year}{2020}) \bibinfo{pages}{222002}.
  \DOIprefix\doi{10.1103/PhysRevLett.125.222002}.
  \href{http://arxiv.org/abs/2005.03016}{{\tt arXiv:2005.03016}}.
\bibitem[{Czakon et~al.(2021)Czakon, Mitov, Pellen, and
  Poncelet}]{Czakon:2020coa}
\bibinfo{author}{M.~Czakon}, \bibinfo{author}{A.~Mitov},
  \bibinfo{author}{M.~Pellen}, \bibinfo{author}{R.~Poncelet},
\newblock \bibinfo{title}{{NNLO QCD predictions for W+c-jet production at the
  LHC}},
\newblock \bibinfo{journal}{JHEP} \bibinfo{volume}{06} (\bibinfo{year}{2021})
  \bibinfo{pages}{100}. \DOIprefix\doi{10.1007/JHEP06(2021)100}.
  \href{http://arxiv.org/abs/2011.01011}{{\tt arXiv:2011.01011}}.
\bibitem[{Gauld et~al.(2023)Gauld, Gehrmann-De~Ridder, Glover, Huss, Garcia,
  and Stagnitto}]{Gauld:2023zlv}
\bibinfo{author}{R.~Gauld}, \bibinfo{author}{A.~Gehrmann-De~Ridder},
  \bibinfo{author}{E.~W.~N. Glover}, \bibinfo{author}{A.~Huss},
  \bibinfo{author}{A.~R. Garcia}, \bibinfo{author}{G.~Stagnitto},
\newblock \bibinfo{title}{{NNLO QCD predictions for Z-boson production in
  association with a charm jet within the LHCb fiducial region}}
  (\bibinfo{year}{2023}). \href{http://arxiv.org/abs/2302.12844}{{\tt
  arXiv:2302.12844}}.
\bibitem[{Aaij et~al.(2022)}]{LHCb:2021stx}
\bibinfo{author}{R.~Aaij}, et~al. (\bibinfo{collaboration}{LHCb}),
\newblock \bibinfo{title}{{Study of Z Bosons Produced in Association with Charm
  in the Forward Region}},
\newblock \bibinfo{journal}{Phys. Rev. Lett.} \bibinfo{volume}{128}
  (\bibinfo{year}{2022}) \bibinfo{pages}{082001}.
  \DOIprefix\doi{10.1103/PhysRevLett.128.082001}.
  \href{http://arxiv.org/abs/2109.08084}{{\tt arXiv:2109.08084}}.
\bibitem[{Banfi et~al.(2006)Banfi, Salam, and Zanderighi}]{Banfi:2006hf}
\bibinfo{author}{A.~Banfi}, \bibinfo{author}{G.~P. Salam},
  \bibinfo{author}{G.~Zanderighi},
\newblock \bibinfo{title}{{Infrared safe definition of jet flavor}},
\newblock \bibinfo{journal}{Eur. Phys. J. C} \bibinfo{volume}{47}
  (\bibinfo{year}{2006}) \bibinfo{pages}{113--124}.
  \DOIprefix\doi{10.1140/epjc/s2006-02552-4}.
  \href{http://arxiv.org/abs/hep-ph/0601139}{{\tt arXiv:hep-ph/0601139}}.
\bibitem[{Czakon et~al.(2022)Czakon, Mitov, and Poncelet}]{Czakon:2022wam}
\bibinfo{author}{M.~Czakon}, \bibinfo{author}{A.~Mitov},
  \bibinfo{author}{R.~Poncelet},
\newblock \bibinfo{title}{{Infrared-safe flavoured anti-$k_T$ jets}}
  (\bibinfo{year}{2022}). \href{http://arxiv.org/abs/2205.11879}{{\tt
  arXiv:2205.11879}}.
\bibitem[{Harris et~al.(1996)Harris, Smith, and Vogt}]{Harris:1995jx}
\bibinfo{author}{B.~W. Harris}, \bibinfo{author}{J.~Smith},
  \bibinfo{author}{R.~Vogt},
\newblock \bibinfo{title}{{Reanalysis of the EMC charm production data with
  extrinsic and intrinsic charm at NLO}},
\newblock \bibinfo{journal}{Nucl. Phys. B} \bibinfo{volume}{461}
  (\bibinfo{year}{1996}) \bibinfo{pages}{181--196}.
  \DOIprefix\doi{10.1016/0550-3213(95)00652-4}.
  \href{http://arxiv.org/abs/hep-ph/9508403}{{\tt arXiv:hep-ph/9508403}}.
\bibitem[{Hou et~al.(2021)}]{Hou:2019efy}
\bibinfo{author}{T.-J. Hou}, et~al.,
\newblock \bibinfo{title}{{New CTEQ global analysis of quantum chromodynamics
  with high-precision data from the LHC}},
\newblock \bibinfo{journal}{Phys. Rev. D} \bibinfo{volume}{103}
  (\bibinfo{year}{2021}) \bibinfo{pages}{014013}.
  \DOIprefix\doi{10.1103/PhysRevD.103.014013}.
  \href{http://arxiv.org/abs/1912.10053}{{\tt arXiv:1912.10053}}.
\bibitem[{Hou et~al.(2022)Hou, Lin, Yan, and Yuan}]{Hou:2022sdf}
\bibinfo{author}{T.-J. Hou}, \bibinfo{author}{H.-W. Lin},
  \bibinfo{author}{M.~Yan}, \bibinfo{author}{C.~P. Yuan},
  \bibinfo{title}{{Impact of lattice $s(x)-\bar{s}(x)$ data in the CTEQ-TEA
  global analysis}}, \bibinfo{year}{2022}.
  \href{http://arxiv.org/abs/2204.07944}{{\tt arXiv:2204.07944}}.
\bibitem[{Blümlein(2016)}]{Blumlein:2015qcn}
\bibinfo{author}{J.~Blümlein},
\newblock \bibinfo{title}{{A Kinematic Condition on Intrinsic Charm}},
\newblock \bibinfo{journal}{Phys. Lett.} \bibinfo{volume}{B753}
  (\bibinfo{year}{2016}) \bibinfo{pages}{619--621}.
  \DOIprefix\doi{10.1016/j.physletb.2015.12.068}.
  \href{http://arxiv.org/abs/1511.00229}{{\tt arXiv:1511.00229}}.
\bibitem[{Bailey et~al.(2021)Bailey, Cridge, Harland-Lang, Martin, and
  Thorne}]{Bailey:2020ooq}
\bibinfo{author}{S.~Bailey}, \bibinfo{author}{T.~Cridge},
  \bibinfo{author}{L.~A. Harland-Lang}, \bibinfo{author}{A.~D. Martin},
  \bibinfo{author}{R.~S. Thorne},
\newblock \bibinfo{title}{{Parton distributions from LHC, HERA, Tevatron and
  fixed target data: MSHT20 PDFs}},
\newblock \bibinfo{journal}{Eur. Phys. J. C} \bibinfo{volume}{81}
  (\bibinfo{year}{2021}) \bibinfo{pages}{341}.
  \DOIprefix\doi{10.1140/epjc/s10052-021-09057-0}.
  \href{http://arxiv.org/abs/2012.04684}{{\tt arXiv:2012.04684}}.
\bibitem[{McGowan et~al.(2022)McGowan, Cridge, Harland-Lang, and
  Thorne}]{McGowan:2022nag}
\bibinfo{author}{J.~McGowan}, \bibinfo{author}{T.~Cridge},
  \bibinfo{author}{L.~A. Harland-Lang}, \bibinfo{author}{R.~S. Thorne},
\newblock \bibinfo{title}{{Approximate N$^{3}$LO Parton Distribution Functions
  with Theoretical Uncertainties: MSHT20aN$^3$LO PDFs}}
  (\bibinfo{year}{2022}). \href{http://arxiv.org/abs/2207.04739}{{\tt
  arXiv:2207.04739}}.
\bibitem[{Gao et~al.(2013)Gao, Guzzi, and Nadolsky}]{Gao:2013wwa}
\bibinfo{author}{J.~Gao}, \bibinfo{author}{M.~Guzzi}, \bibinfo{author}{P.~M.
  Nadolsky},
\newblock \bibinfo{title}{{Charm quark mass dependence in a global QCD
  analysis}},
\newblock \bibinfo{journal}{Eur. Phys. J.} \bibinfo{volume}{C73}
  (\bibinfo{year}{2013}) \bibinfo{pages}{2541}.
  \DOIprefix\doi{10.1140/epjc/s10052-013-2541-4}.
  \href{http://arxiv.org/abs/1304.3494}{{\tt arXiv:1304.3494}}.
\bibitem[{Courtoy et~al.(2023)Courtoy, Huston, Nadolsky, Xie, Yan, and
  Yuan}]{Courtoy:2022ocu}
\bibinfo{author}{A.~Courtoy}, \bibinfo{author}{J.~Huston},
  \bibinfo{author}{P.~Nadolsky}, \bibinfo{author}{K.~Xie},
  \bibinfo{author}{M.~Yan}, \bibinfo{author}{C.~P. Yuan},
\newblock \bibinfo{title}{{Parton distributions need representative sampling}},
\newblock \bibinfo{journal}{Phys. Rev. D} \bibinfo{volume}{107}
  (\bibinfo{year}{2023}) \bibinfo{pages}{034008}.
  \DOIprefix\doi{10.1103/PhysRevD.107.034008}.
  \href{http://arxiv.org/abs/2205.10444}{{\tt arXiv:2205.10444}}.
\bibitem[{Aivazis et~al.(1994)Aivazis, Collins, Olness, and
  Tung}]{Aivazis:1993pi}
\bibinfo{author}{M.~A.~G. Aivazis}, \bibinfo{author}{J.~C. Collins},
  \bibinfo{author}{F.~I. Olness}, \bibinfo{author}{W.-K. Tung},
\newblock \bibinfo{title}{{Leptoproduction of heavy quarks. 2. A Unified QCD
  formulation of charged and neutral current processes from fixed target to
  collider energies}},
\newblock \bibinfo{journal}{Phys. Rev.} \bibinfo{volume}{D50}
  (\bibinfo{year}{1994}) \bibinfo{pages}{3102--3118}.
  \DOIprefix\doi{10.1103/PhysRevD.50.3102}.
  \href{http://arxiv.org/abs/hep-ph/9312319}{{\tt arXiv:hep-ph/9312319}}.
\bibitem[{Guzzi et~al.(2012)Guzzi, Nadolsky, Lai, and Yuan}]{Guzzi:2011ew}
\bibinfo{author}{M.~Guzzi}, \bibinfo{author}{P.~M. Nadolsky},
  \bibinfo{author}{H.-L. Lai}, \bibinfo{author}{C.~P. Yuan},
\newblock \bibinfo{title}{{General-Mass Treatment for Deep Inelastic Scattering
  at Two-Loop Accuracy}},
\newblock \bibinfo{journal}{Phys. Rev.} \bibinfo{volume}{D86}
  (\bibinfo{year}{2012}) \bibinfo{pages}{053005}.
  \DOIprefix\doi{10.1103/PhysRevD.86.053005}.
  \href{http://arxiv.org/abs/1108.5112}{{\tt arXiv:1108.5112}}.
\bibitem[{Guzzi et~al.(2011)Guzzi, Nadolsky, and Olness}]{GuzziIC:2011}
\bibinfo{author}{M.~Guzzi}, \bibinfo{author}{P.~Nadolsky},
  \bibinfo{author}{F.~Olness},
\newblock \bibinfo{title}{\protect{Probing intrinsic charm at the EIC}},
\newblock in: \bibinfo{booktitle}{{Gluons and the quark sea at high energies:
  Distributions, polarization, tomography}}, \bibinfo{year}{2011},
  p.~\bibinfo{pages}{44}. \href{http://arxiv.org/abs/1108.1713}{{\tt
  arXiv:1108.1713}}.
\bibitem[{Anchordoqui et~al.(2022)}]{Anchordoqui:2021ghd}
\bibinfo{author}{L.~A. Anchordoqui}, et~al.,
\newblock \bibinfo{title}{{The Forward Physics Facility: Sites, experiments,
  and physics potential}},
\newblock \bibinfo{journal}{Phys. Rept.} \bibinfo{volume}{968}
  (\bibinfo{year}{2022}) \bibinfo{pages}{1--50}.
  \DOIprefix\doi{10.1016/j.physrep.2022.04.004}.
  \href{http://arxiv.org/abs/2109.10905}{{\tt arXiv:2109.10905}}.
\bibitem[{Feng et~al.(2022)}]{Feng:2022inv}
\bibinfo{author}{J.~L. Feng}, et~al., \bibinfo{title}{{The Forward Physics
  Facility at the High-Luminosity LHC}}, \bibinfo{year}{2022}.
  \href{http://arxiv.org/abs/2203.05090}{{\tt arXiv:2203.05090}}.
\bibitem[{Sufian et~al.(2020)Sufian, Liu, Alexandru, Brodsky, de~T\'eramond,
  Dosch, Draper, Liu, and Yang}]{Sufian:2020coz}
\bibinfo{author}{R.~S. Sufian}, \bibinfo{author}{T.~Liu},
  \bibinfo{author}{A.~Alexandru}, \bibinfo{author}{S.~J. Brodsky},
  \bibinfo{author}{G.~F. de~T\'eramond}, \bibinfo{author}{H.~G. Dosch},
  \bibinfo{author}{T.~Draper}, \bibinfo{author}{K.-F. Liu},
  \bibinfo{author}{Y.-B. Yang},
\newblock \bibinfo{title}{{Constraints on charm-anticharm asymmetry in the
  nucleon from lattice QCD}},
\newblock \bibinfo{journal}{Phys. Lett. B} \bibinfo{volume}{808}
  (\bibinfo{year}{2020}) \bibinfo{pages}{135633}.
  \DOIprefix\doi{10.1016/j.physletb.2020.135633}.
  \href{http://arxiv.org/abs/2003.01078}{{\tt arXiv:2003.01078}}.
\bibitem[{Ji(2013)}]{Ji:2013dva}
\bibinfo{author}{X.~Ji},
\newblock \bibinfo{title}{{Parton Physics on a Euclidean Lattice}},
\newblock \bibinfo{journal}{Phys. Rev. Lett.} \bibinfo{volume}{110}
  (\bibinfo{year}{2013}) \bibinfo{pages}{262002}.
  \DOIprefix\doi{10.1103/PhysRevLett.110.262002}.
  \href{http://arxiv.org/abs/1305.1539}{{\tt arXiv:1305.1539}}.
\bibitem[{Radyushkin(2017)}]{Radyushkin:2017cyf}
\bibinfo{author}{A.~V. Radyushkin},
\newblock \bibinfo{title}{{Quasi-parton distribution functions, momentum
  distributions, and pseudo-parton distribution functions}},
\newblock \bibinfo{journal}{Phys. Rev. D} \bibinfo{volume}{96}
  (\bibinfo{year}{2017}) \bibinfo{pages}{034025}.
  \DOIprefix\doi{10.1103/PhysRevD.96.034025}.
  \href{http://arxiv.org/abs/1705.01488}{{\tt arXiv:1705.01488}}.
\bibitem[{Zhang et~al.(2021)Zhang, Lin, and Yoon}]{Zhang:2020dkn}
\bibinfo{author}{R.~Zhang}, \bibinfo{author}{H.-W. Lin},
  \bibinfo{author}{B.~Yoon},
\newblock \bibinfo{title}{{Probing nucleon strange and charm distributions with
  lattice QCD}},
\newblock \bibinfo{journal}{Phys. Rev. D} \bibinfo{volume}{104}
  (\bibinfo{year}{2021}) \bibinfo{pages}{094511}.
  \DOIprefix\doi{10.1103/PhysRevD.104.094511}.
  \href{http://arxiv.org/abs/2005.01124}{{\tt arXiv:2005.01124}}.
\bibitem[{Campbell et~al.(2004)Campbell, Ellis, Maltoni, and
  Willenbrock}]{Campbell:2003dd}
\bibinfo{author}{J.~M. Campbell}, \bibinfo{author}{R.~K. Ellis},
  \bibinfo{author}{F.~Maltoni}, \bibinfo{author}{S.~Willenbrock},
\newblock \bibinfo{title}{{Associated production of a $Z$ Boson and a single
  heavy quark jet}},
\newblock \bibinfo{journal}{Phys. Rev. D} \bibinfo{volume}{69}
  (\bibinfo{year}{2004}) \bibinfo{pages}{074021}.
  \DOIprefix\doi{10.1103/PhysRevD.69.074021}.
  \href{http://arxiv.org/abs/hep-ph/0312024}{{\tt arXiv:hep-ph/0312024}}.
\bibitem[{Butterworth et~al.(2016)}]{Butterworth:2015oua}
\bibinfo{author}{J.~Butterworth}, et~al.,
\newblock \bibinfo{title}{{PDF4LHC recommendations for LHC Run II}},
\newblock \bibinfo{journal}{J. Phys.} \bibinfo{volume}{G43}
  (\bibinfo{year}{2016}) \bibinfo{pages}{023001}.
  \DOIprefix\doi{10.1088/0954-3899/43/2/023001}.
  \href{http://arxiv.org/abs/1510.03865}{{\tt arXiv:1510.03865}}.
\bibitem[{Ball et~al.(2022)}]{NNPDF:2021njg}
\bibinfo{author}{R.~D. Ball}, et~al. (\bibinfo{collaboration}{NNPDF}),
\newblock \bibinfo{title}{{The path to proton structure at 1\% accuracy}},
\newblock \bibinfo{journal}{Eur. Phys. J. C} \bibinfo{volume}{82}
  (\bibinfo{year}{2022}) \bibinfo{pages}{428}.
  \DOIprefix\doi{10.1140/epjc/s10052-022-10328-7}.
  \href{http://arxiv.org/abs/2109.02653}{{\tt arXiv:2109.02653}}.
\bibitem[{CMS Collaboration(2016)}]{CMS:2016dyh}
CMS Collaboration, \bibinfo{title}{{Measurement of associated Z + charm
  production in pp collisions at $\sqrt{s} = 8~\mathrm{TeV}$}},
  \bibinfo{howpublished}{CMS-PAS-SMP-15-009}, \bibinfo{year}{2016}.
\bibitem[{Gao and Nadolsky(2014)}]{Gao:2013bia}
\bibinfo{author}{J.~Gao}, \bibinfo{author}{P.~Nadolsky},
\newblock \bibinfo{title}{{A meta-analysis of parton distribution functions}},
\newblock \bibinfo{journal}{JHEP} \bibinfo{volume}{07} (\bibinfo{year}{2014})
  \bibinfo{pages}{035}. \DOIprefix\doi{10.1007/JHEP07(2014)035}.
  \href{http://arxiv.org/abs/1401.0013}{{\tt arXiv:1401.0013}}.

\end{thebibliography}

\clearpage \newpage

\appendix
\setcounter{figure}{0}
\setcounter{page}{1}
\section{\bf Supplementary discussion}

\newcommand{\xFC}{\langle x\rangle_\text{FC}}
\subsection*{\bf FC momentum fractions from Lagrange Multiplier scans}
\begin{figure}[htb]
\center
\includegraphics[width=0.48\textwidth]{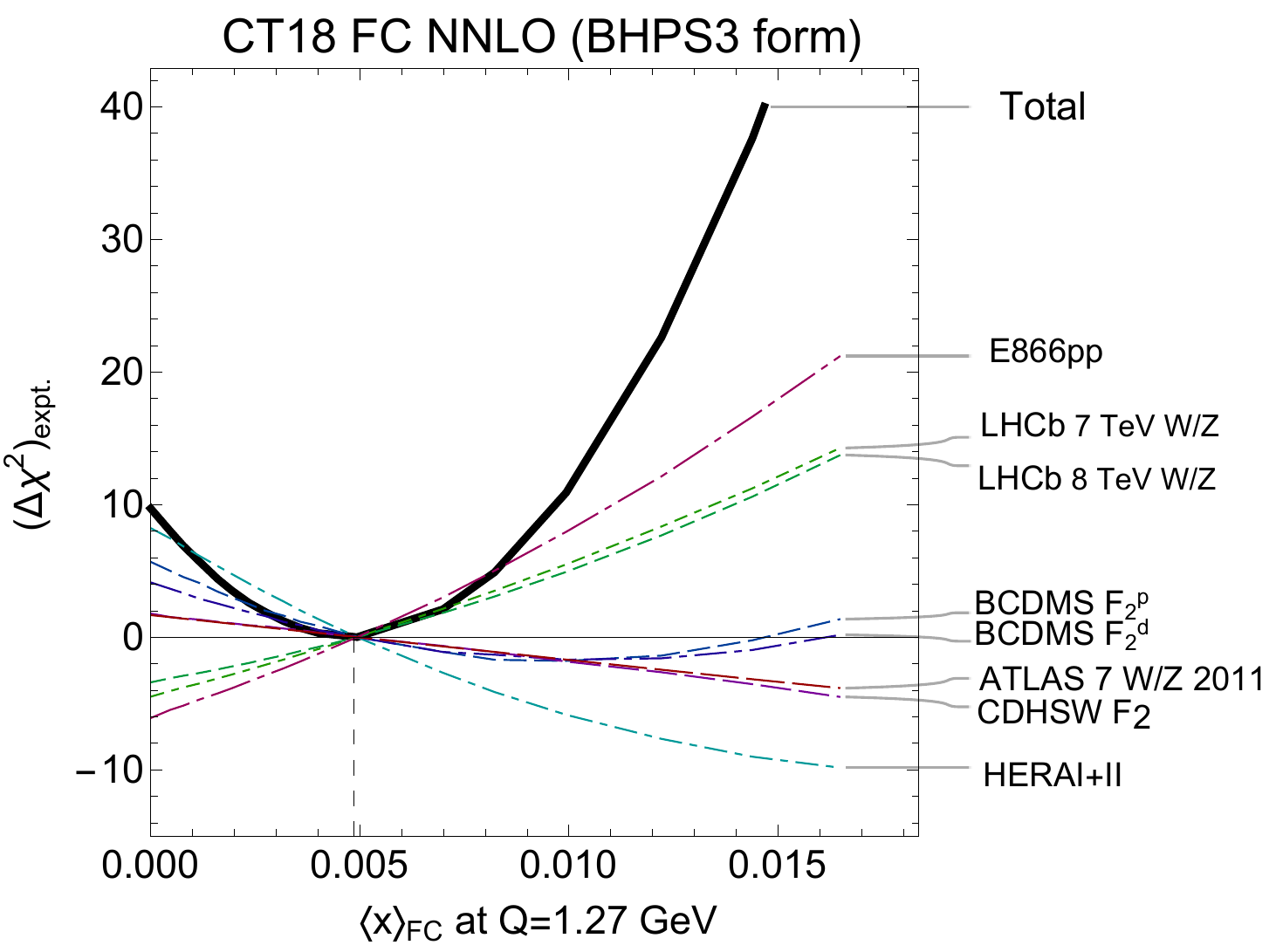}
\includegraphics[width=0.48\textwidth]{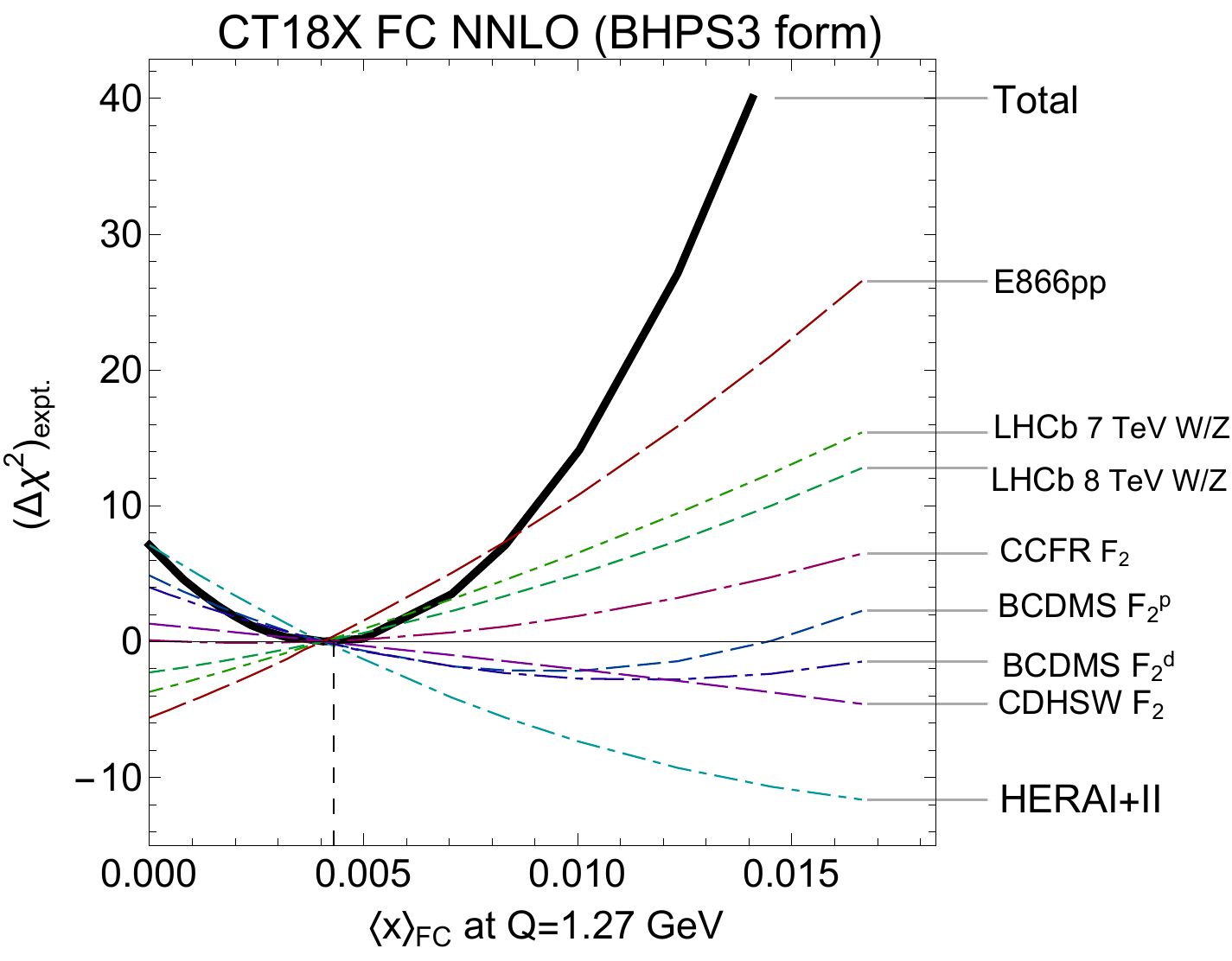}
\includegraphics[width=0.48\textwidth]{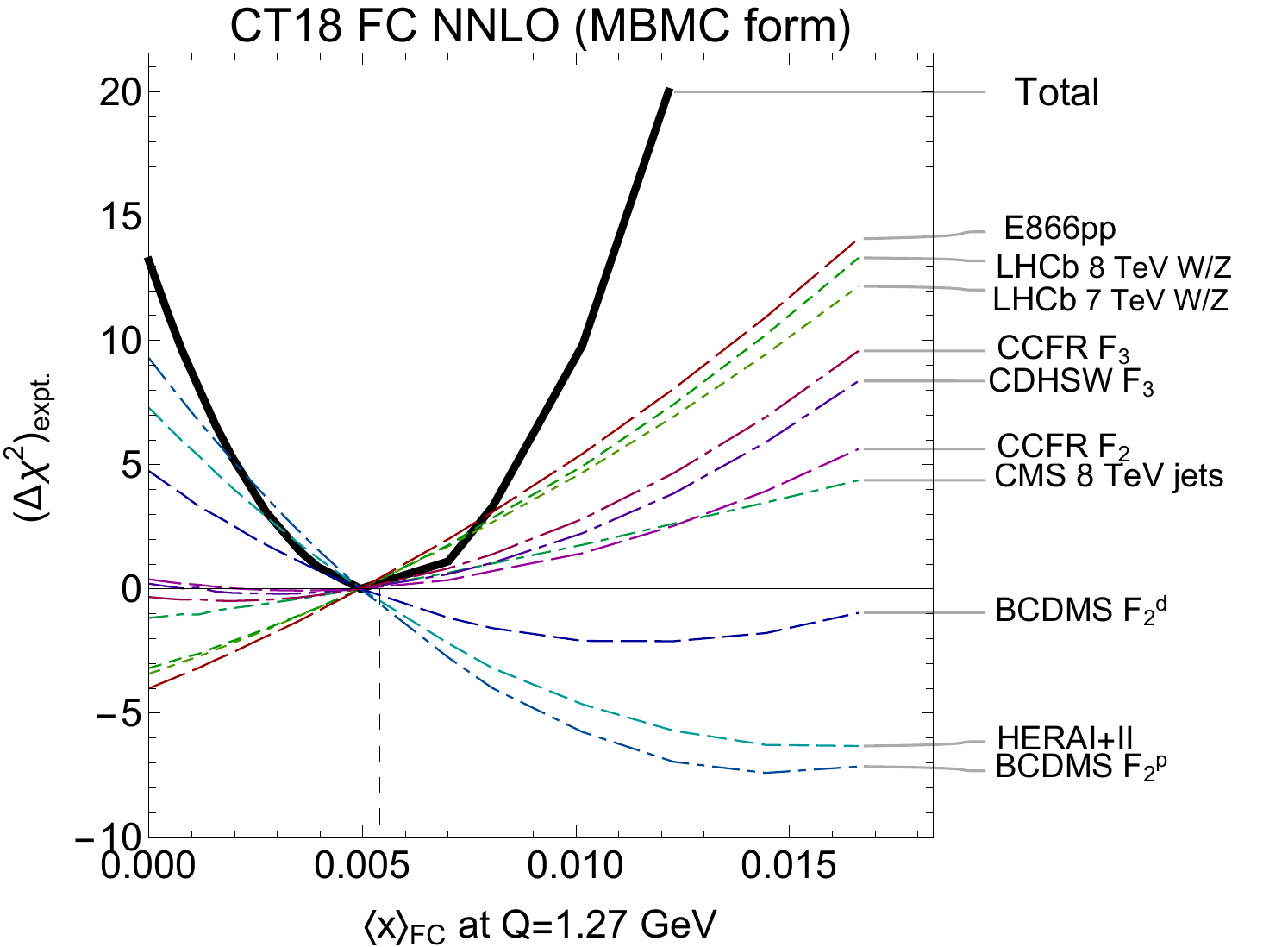}
\includegraphics[width=0.48\textwidth]{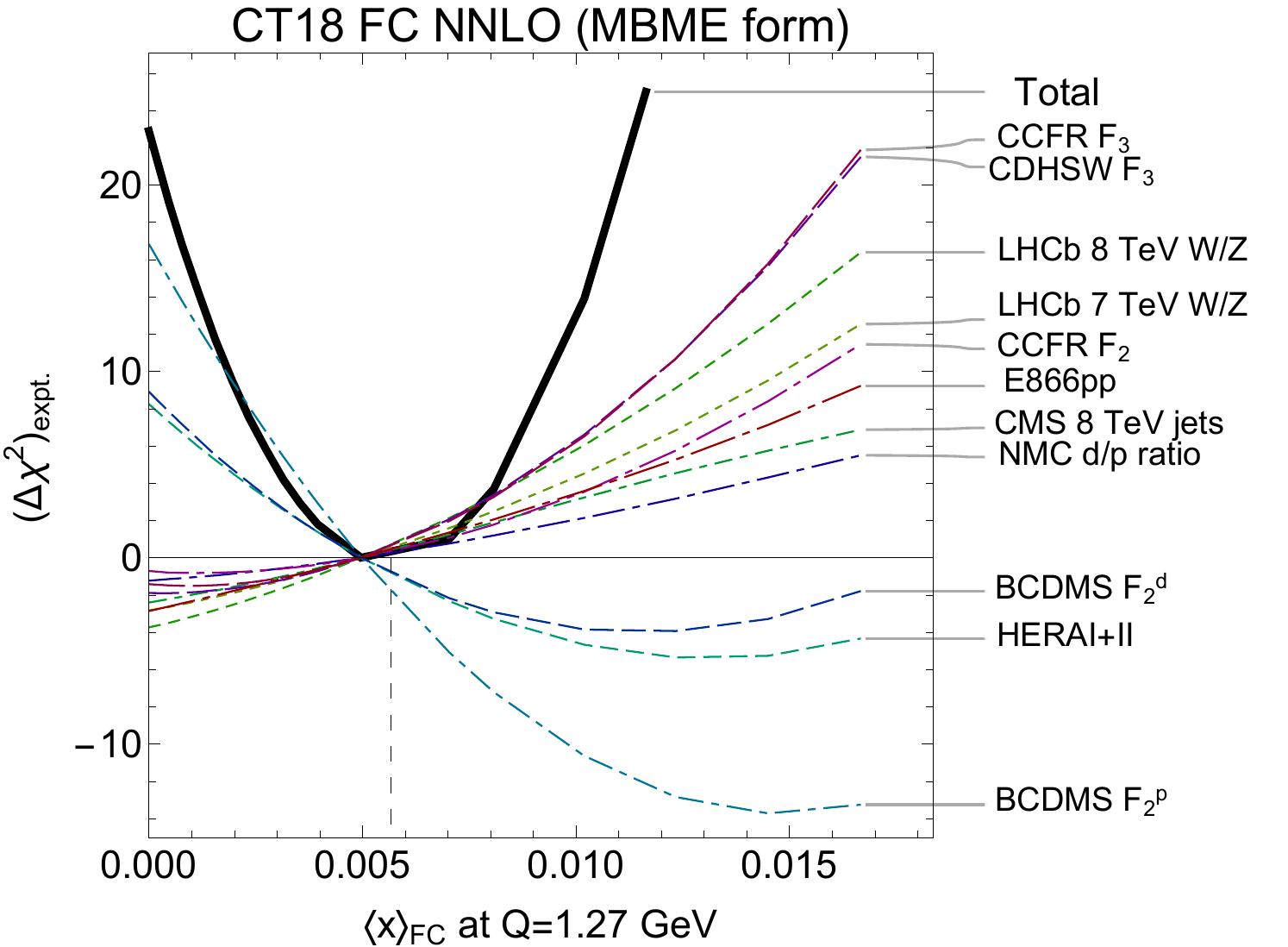}
\caption{Lagrange Multiplier scans on $\xFC$ for various FC scenarios.}
\label{fig:FC-chi2-LM}
\end{figure}

In Fig.~\ref{fig:FC-chi2-LM}, we illustrate the $\chi^2$ dependence on the total momentum fraction, $\xFC \equiv \langle x \rangle_{c^+} (Q=Q_0)$, carried by fitted charm (FC) for each of the scenarios examined in the main paper; for this, we use the Lagrange multiplier (LM) method adopted in the CT18 NNLO analysis~\cite{Hou:2019efy} and related publications. The LM scans identify the most constraining experiments whose $\chi^2$ depends strongly on $\xFC$. In all cases, we observe that the preferred range of $\xFC$ is determined by trade-offs between the pulls of the largest NC DIS data sets from the combined HERA and BCDMS measurements, which prefer larger $\xFC$, and vector-boson production cross sections from E866 and LHCb $pp$ scattering, which prefer lower $\xFC$. For the meson-baryon models (MBMs), we also observe more pronounced downward pulls from the CC DIS ($F_3$) measurements by CCFR and CDHSW, as well as some pulls from the CMS 8 TeV jet production and NMC $d/p$ DIS ratio. The structure function $F_3$ depends on $(s\! -\! {\bar s})$ and  $(c\! -\! {\bar c})$ already at leading order; for this reason, measurements of this quantity may possess more nominal sensitivity to the $c\! \neq\! \bar{c}$ asymmetry which tracks $\langle x \rangle_\mathrm{FC}$ in the MBM as reflected by the CCFR and CDHSW curves in the lower panels of Fig.~\ref{fig:FC-chi2-LM}.

The strengths of the pulls depend on the settings used in the associated analysis, as can be seen, {\it e.g.}, from the comparison of the pulls in the CT18 and CT18X NNLO fits with the BHPS3 model in the upper row of Fig.~\ref{fig:FC-chi2-LM}. For all examined models, $\xFC \approx 0.5\%$ is preferred even more mildly than in the CT14 IC study. [In the latter, the BHPS3 fitted charm with $\xFC \approx 1\%$ improved $\chi^2$ by nearly 40 units. In the CT18 BHPS3 analysis, the analogous improvement is $\sim\! 10$ units.]

From Fig.~\ref{fig:FC-chi2-LM}, which also corresponds to Fig.~2 (left) of the main paper, we read off the allowed ranges for the moment $\xFC$:
\begin{eqnarray}
 \xFC  = &  0.0048 {}^{+0.0063}_{-0.0043}\  ({}^{+0.0090}_{-0.0048})  & \mbox{  for CT18 NNLO (BHPS3 form);} \nonumber \\
 \xFC  = & 0.0041 {}^{+0.0049}_{-0.0041}\  ({}^{+0.0091}_{-0.0041})  & \mbox{  for CT18X NNLO (BHPS3 form);} \nonumber \\
 \xFC  = & 0.0057 {}^{+0.0048}_{-0.0045}\  ({}^{+0.0084}_{-0.0057})  & \mbox{  for CT18 NNLO (MBMC form);} 
\nonumber \\
 \xFC  = & 0.0061 {}^{+0.0030}_{-0.0038}\  ({}^{+0.0064}_{-0.0061})  & \mbox{  for CT18 NNLO (MBME form)},
\label{FC-moments}
\end{eqnarray}
where the first and second uncertainties represent the $\Delta \chi^2=10$ and 30 intervals, respectively.

\begin{figure}[tb]\centering
\includegraphics[width=0.49\textwidth]{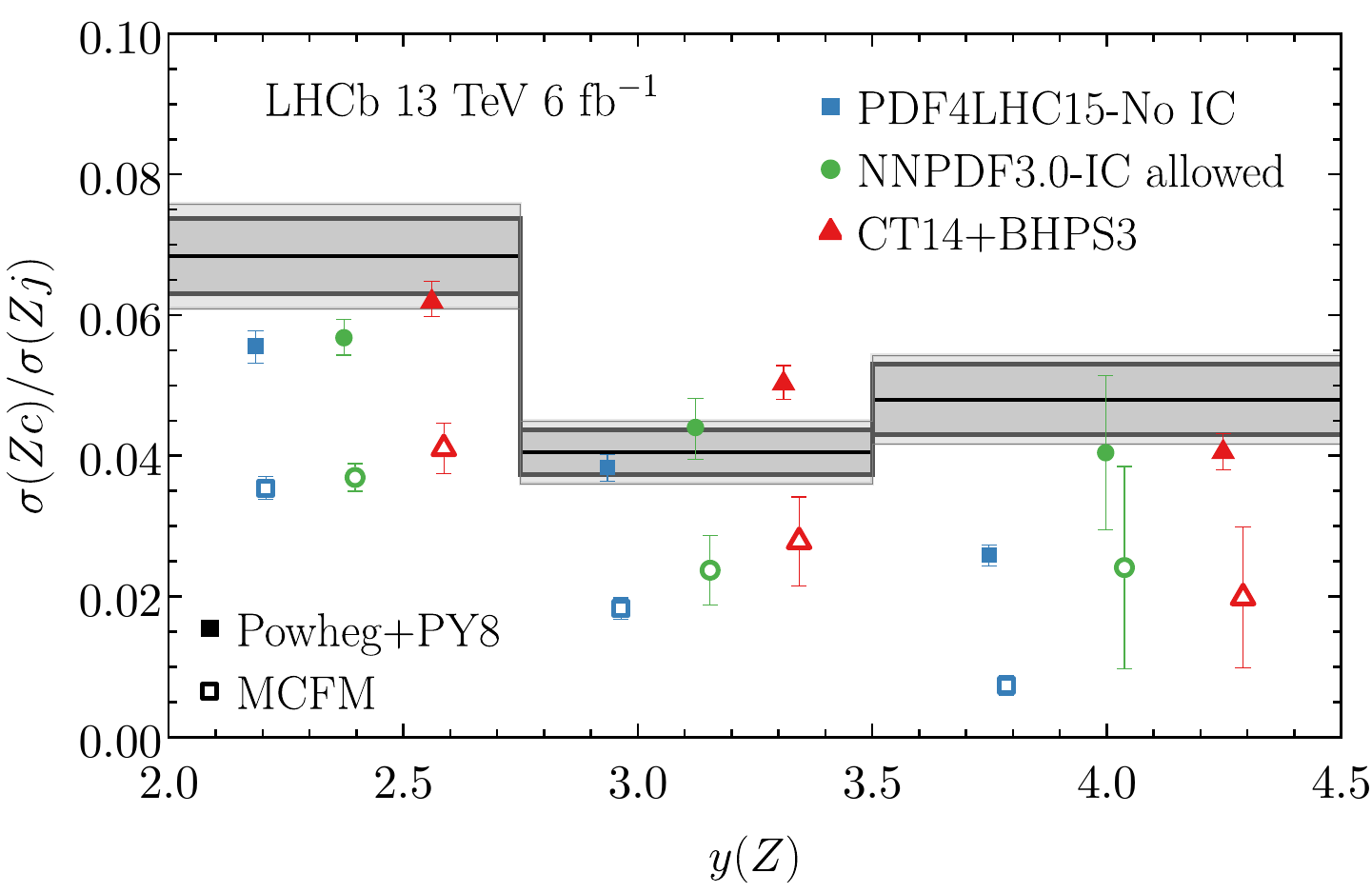}
\includegraphics[width=0.49\textwidth]{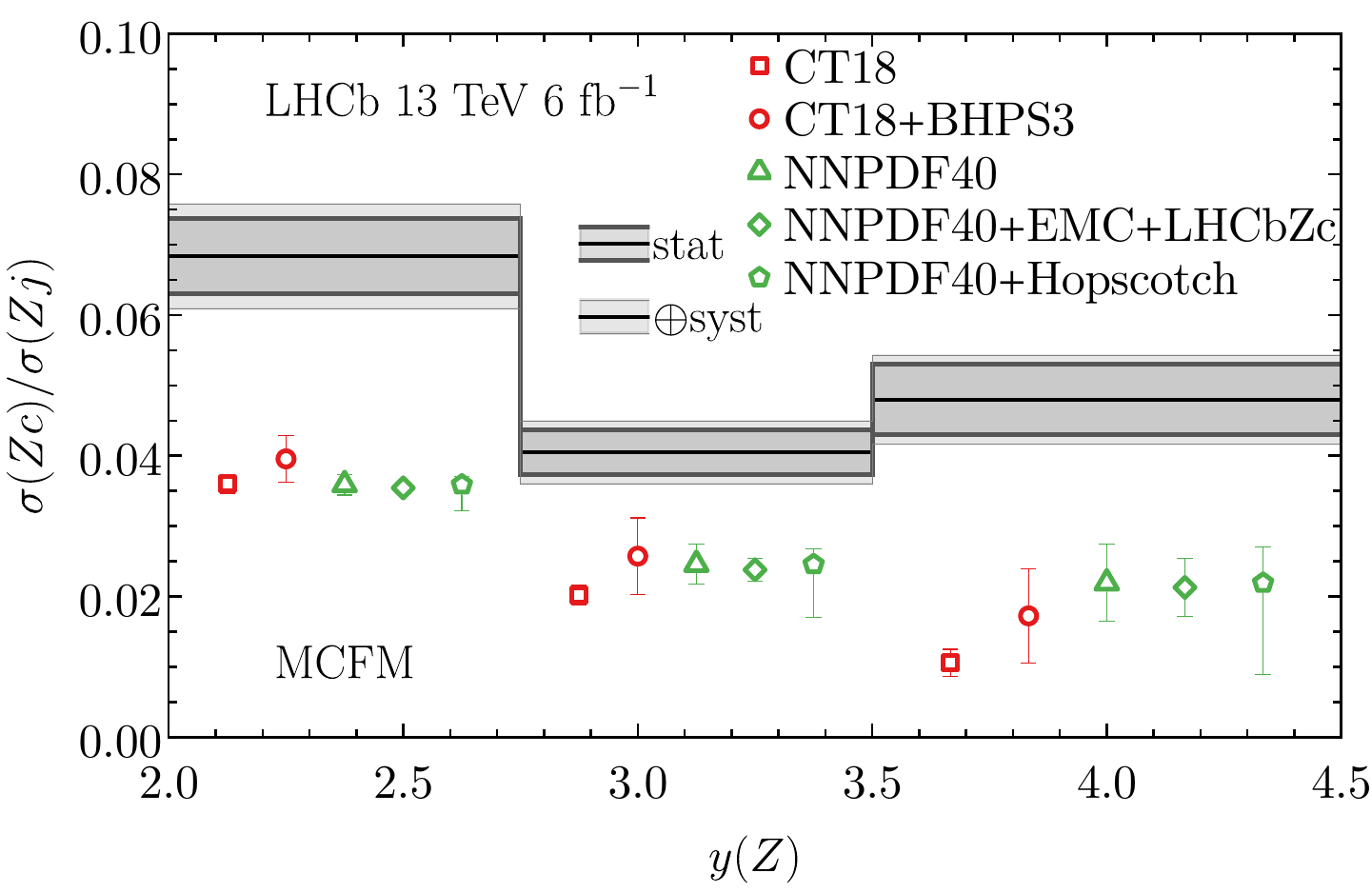}
\caption{Theoretical predictions for the $\sigma(Zc)/\sigma(Zj)$ ratios as a function of $Z$-boson rapidity, $y(Z)$, compared against the recent 13 TeV LHCb measurements~\cite{LHCb:2021stx} (shown as grey bands). Here, the error bars of the MCFM NLO prediction include only the PDF uncertainty.}
\label{fig:LHCb13Rjc}
\end{figure}

\subsection*{\bf PDF uncertainties for the LHCb 13 TeV $\sigma(Zc)/\sigma(Zj)$ measurement}

LHCb has recently measured $Z+c$ production at 13 TeV, with data normalized to the corresponding $Z+\textrm{jet}$ cross sections \cite{LHCb:2021stx}. 
As noted in the main text, the associated production of a $Z$ boson and charm quark, $Z+c$, has been suggested~\cite{Boettcher:2015sqn} as providing a direct probe of the charm PDF, due to the leading contribution from $gc\to Zc$ scattering at parton level.
In this work, we find that any FC contribution to $Z+c$ production in the central region cannot be distinguished from the non-FC PDF uncertainty. In contrast, forward $Z+c$ production may in principle possess more direct sensitivity due
to the Born-level relation between the measured rapidity and probed momentum fraction, $x\sim(Q/\sqrt{s})e^{y}$, such that PDFs in the large-$x$ region, where typical IC models predict the greatest impact, might be constrained.

In Fig.~\ref{fig:LHCb13Rjc}, in addition to the LHCb $Zc/Zj$ ratio data, we show several predictions to illustrate the underlying physics issues.
\begin{enumerate}
    \item The NLO calculation using MCFM \cite{Campbell:2003dd} (indicated by corresponding open symbols), is presented in both panels of Fig.~\ref{fig:LHCb13Rjc} for an array of recent PDF sets without or with the FC component. Error bars on these predictions indicate the PDF uncertainty to gauge its magnitude compared to the other contributing uncertainties.
    \item In the left panel only, we also include predictions made with POWHEG-BOX interfaced to Pythia8~\cite{Boettcher:2015sqn} (indicated by filled symbols) for the indicated PDF sets. The NLO+PS calculation has a significant theoretical uncertainty. To discriminate among the FC models, the theoretical uncertainty of predictions must be smaller than the FC dependence. 
\end{enumerate}

Now turning to the details, in Fig.~\ref{fig:LHCb13Rjc} (left)
we include predictions based on several PDFs: PDF4LHC15 (no FC)~\cite{Butterworth:2015oua}; NNPDF3.0 with FC allowed~\cite{Ball:2016neh}; and CT14 with the BHPS3 IC model~\cite{Hou:2017khm}.  
Here, we notice that parton showering significantly enhances the ratio $\sigma(Zc)/\sigma(Zj)$, reflects the important contribution from higher-order splittings $g\to c\bar{c}$ that are not included at fixed NLO. This enhancement has a sizeable PDF dependence that cannot be captured by a universal $K$-factor approach applied to all regions of phase space. 
The uncertainty of the CT14+BHPS3 PDF fixed-order prediction is larger than that of the respective NLO+PS prediction, simply because the latter prediction, taken from  Ref.~\cite{Boettcher:2015sqn}, does not include the PDF uncertainty. 

Figure~\ref{fig:LHCb13Rjc} (right) shows the MCFM fixed-order NLO calculations with a number of more recent PDF ensembles: CT18~\cite{Hou:2019efy}; CT18 with the BHPS3 FC model presented in the main paper; NNPDF4.0~\cite{NNPDF:2021njg}; NNPDF4.0 additionally fitted to the EMC and LHCb data~\cite{Ball:2022qks}; and the enlarged hopscotch-sampled set of Ref.~\cite{Courtoy:2022ocu}. Here, we wish to gauge the effect of FC before parton showering is included.  
Again, we see the enhancement of the $\sigma(Zc)/\sigma(Zj)$ ratio from the inclusion of FC when comparing the CT18 and CT18+BHPS3 predictions.  
The (nominal) NNPDF4.0 predictions give a slightly larger result than CT18, with a smaller PDF uncertainty. Upon incorporating the LHCb data (along with the EMC $F^c_2$ measurements), the NNPDF4.0 error band slightly shrinks, with the central value largely unchanged. As pointed out in the hopscotch scan study of Ref.~\cite{Courtoy:2022ocu}, the NNPDF PDF uncertainty can be undersampled. Hence, we also compare to a prediction which includes an additional contribution to the NNPDF4.0 uncertainty from the hopscotch-sampled PDF solutions. As expected, this enlarges the PDF uncertainty and makes it closer to that of CT18 BHPS3 prediction.

As a final remark, we observe that the NLO and NLO+PS calculations presented here are not yet sufficient for the extraction of the FC from the LHC data. For the fiducial central region of CMS \cite{CMS:2016dyh}, Ref.~\cite{Hou:2017khm} examined multi-particle final-state effects using the matrix-element plus parton shower merging (MEPS) approach. It found that the sensitivity to FC could be diluted due to additional contributions from $Z+\textrm{non-}c$ partonic subprocesses involving $g\to c\bar{c}$ splitting in the final state. These dilution effects can be even more severe in the forward region measured by the LHCb experiment, as a result of the larger showering effects expected for the associated rapidities. Indeed, in Fig.~5 of \cite{Gauld:2023zlv}, the two NLO+PS predictions based on \texttt{Pythia 8} and \texttt{Herwig 7} bracket the NNLO predictions, and both have twice as large uncertainties as at NNLO.
Correctly combining the final-state gluon splitting ($g \to c\bar{c}$) contribution, determined via parton showers, with the hard part of the NNLO calculation is expected to be challenging.
Furthermore, multiparton interactions  are present in the actual LHCb measurement and modify the NLO+PS predictions in a pattern which can mimic and potentially complicate the extraction of the FC signature (Appendix B in \cite{Gauld:2023zlv}).


\subsection*{\bf Correlation between PDFs and LHCb 13 TeV $Z+c,~Z+\text{jet}$ cross sections}
\begin{figure}[p]
    \centering
    \includegraphics[width=0.32\textwidth]{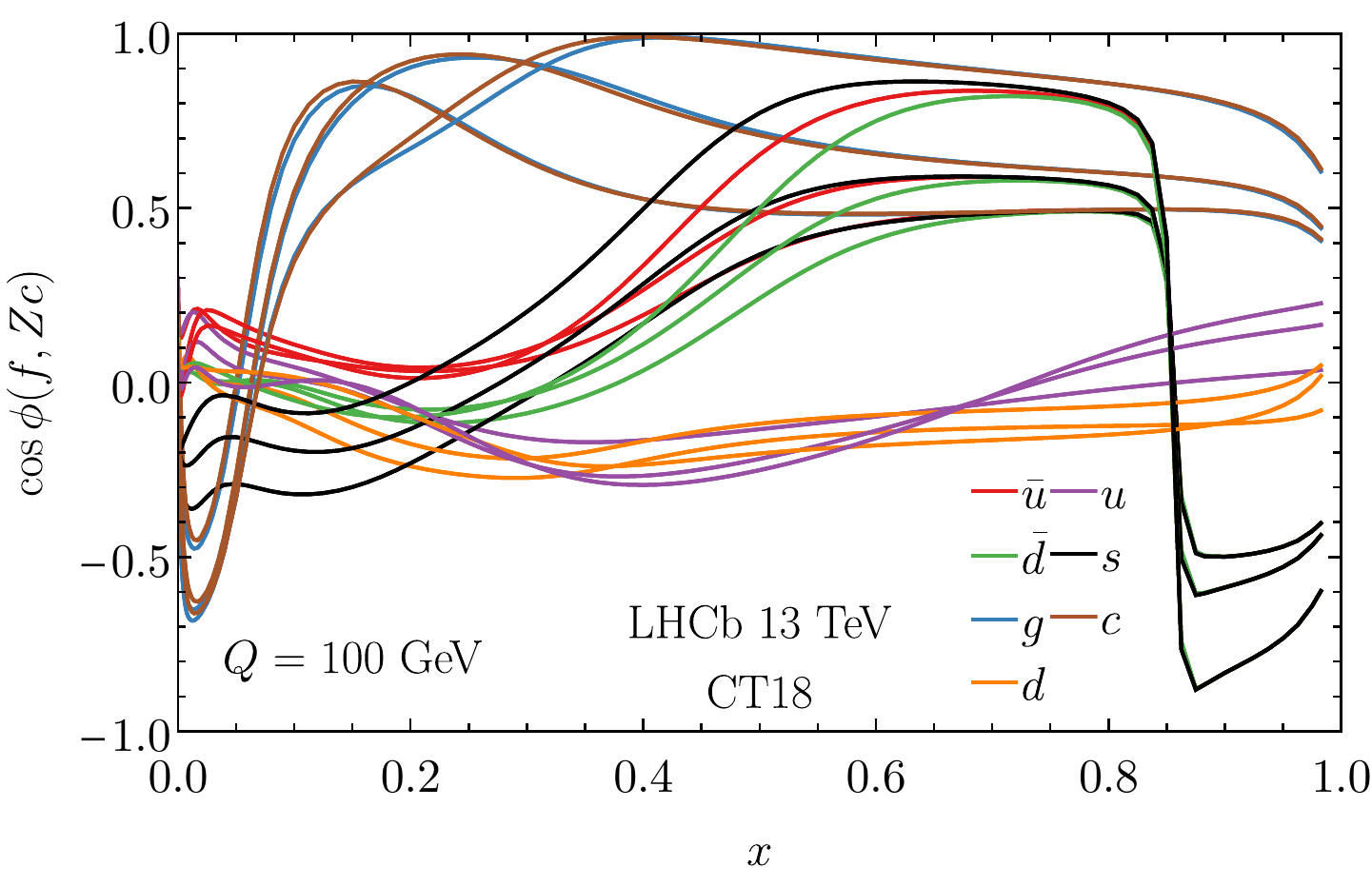}
    \includegraphics[width=0.32\textwidth]{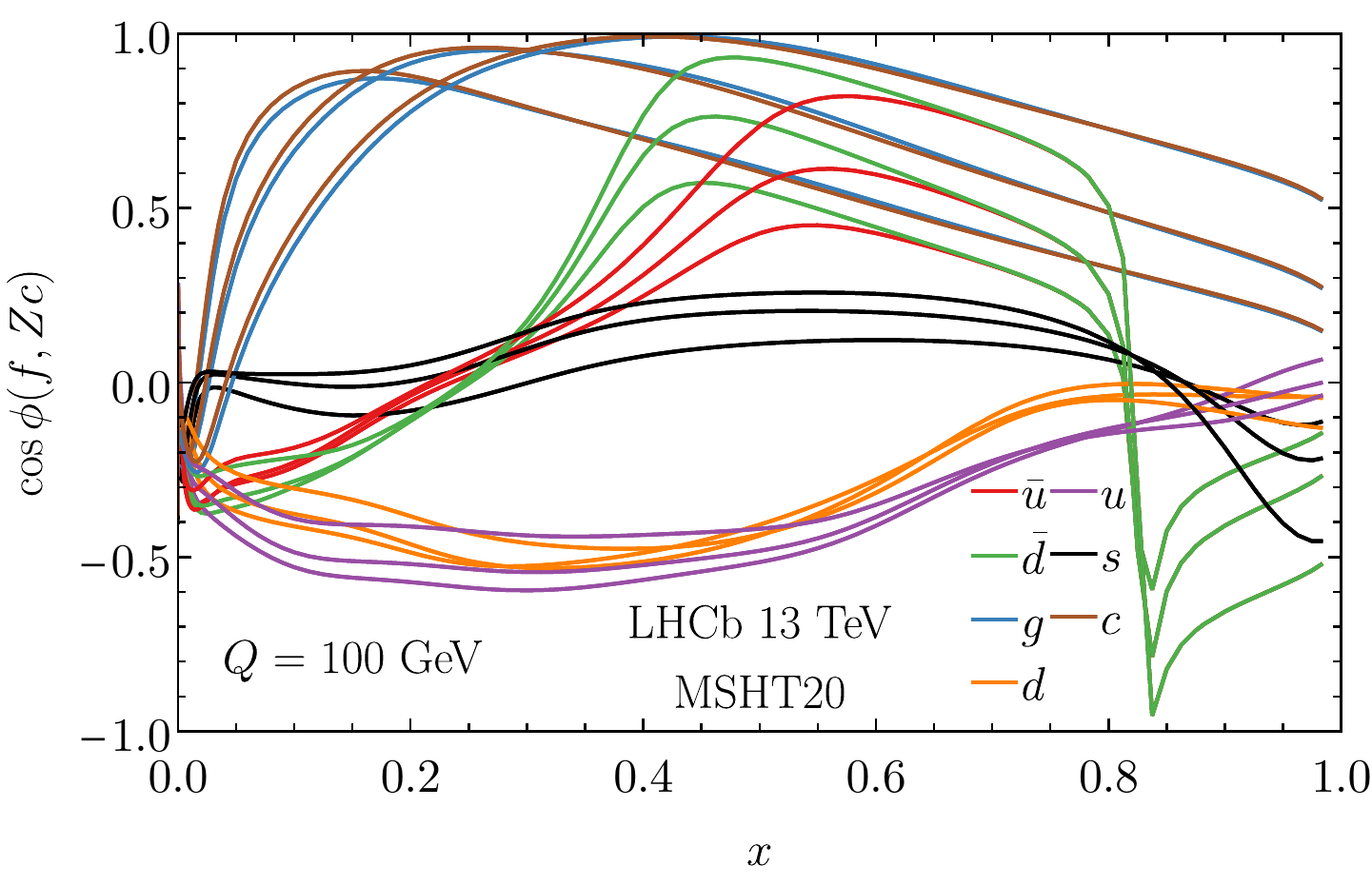} 
    \includegraphics[width=0.32\textwidth]{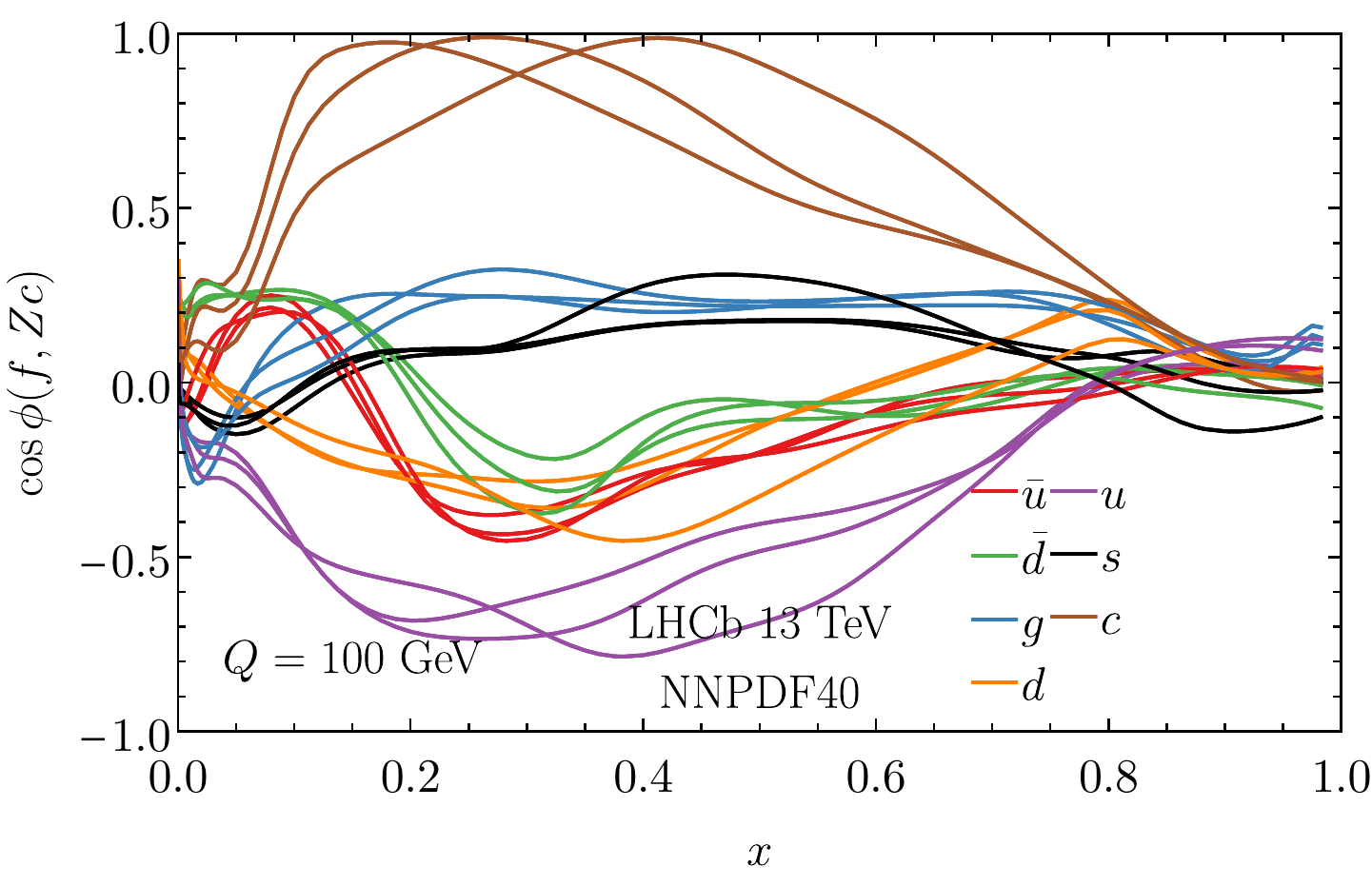}    
    \includegraphics[width=0.32\textwidth]{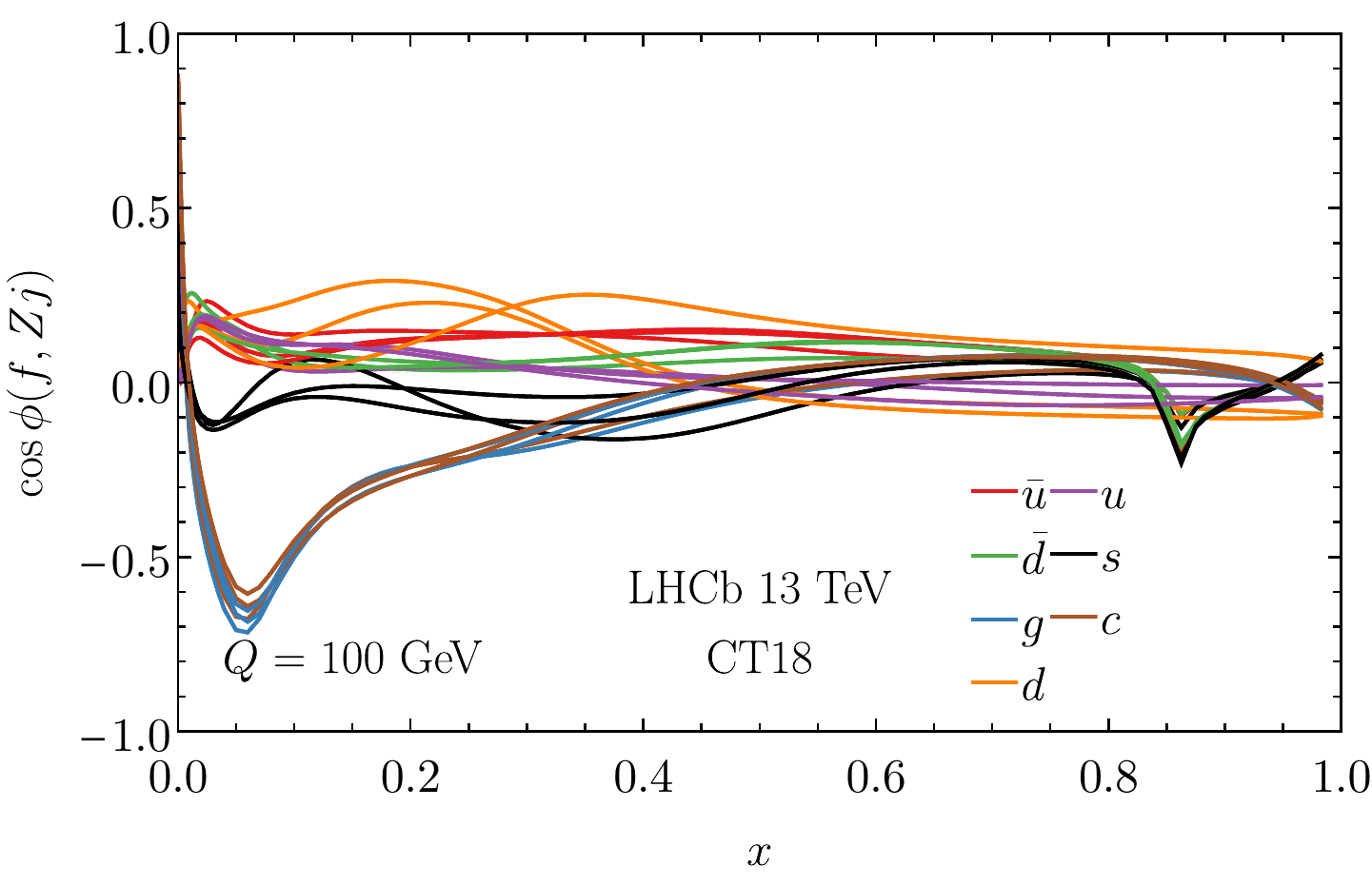}
    \includegraphics[width=0.33\textwidth]{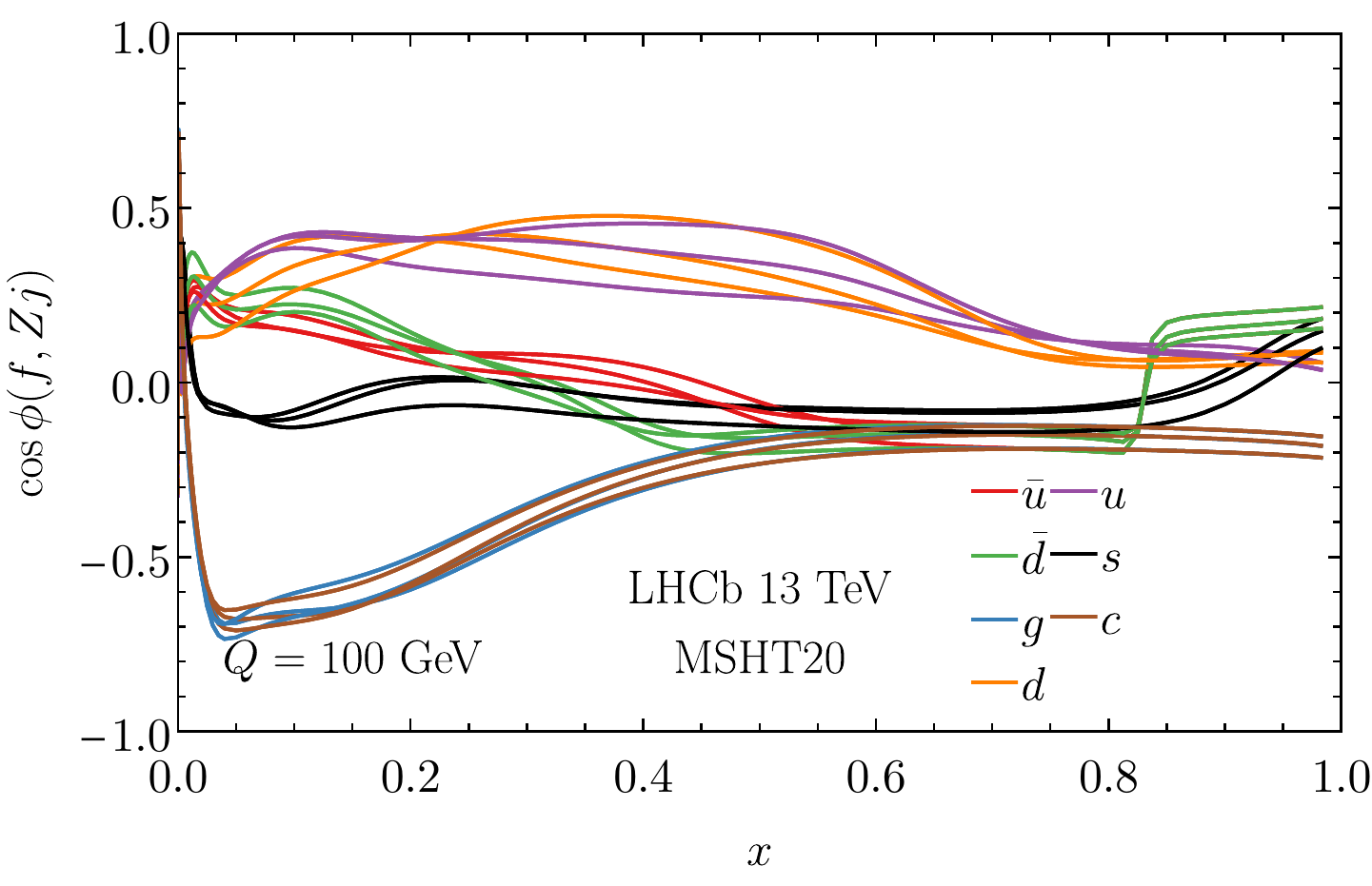}
    \includegraphics[width=0.32\textwidth]{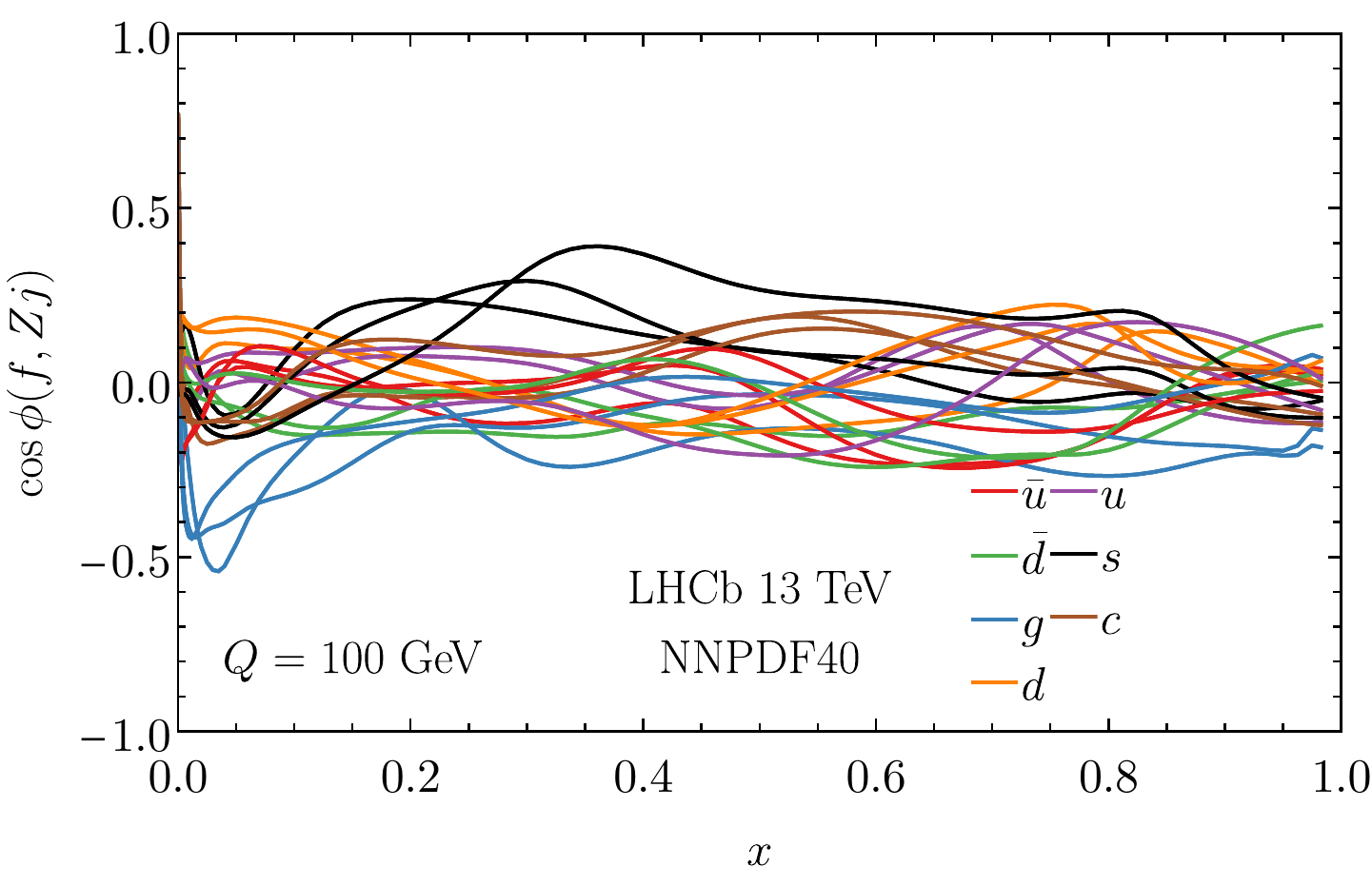}    
    \includegraphics[width=0.32\textwidth]{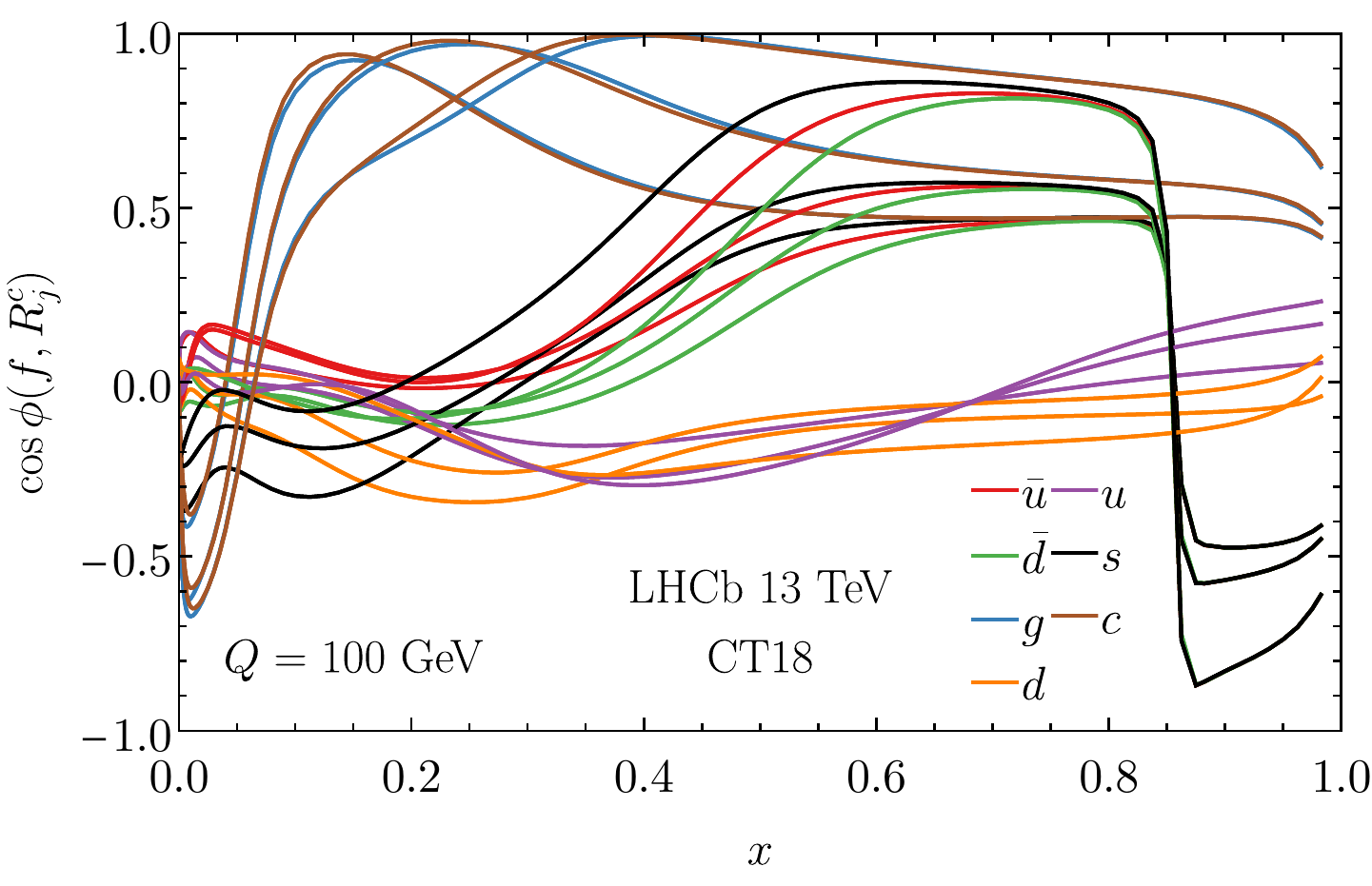}
    \includegraphics[width=0.32\textwidth]{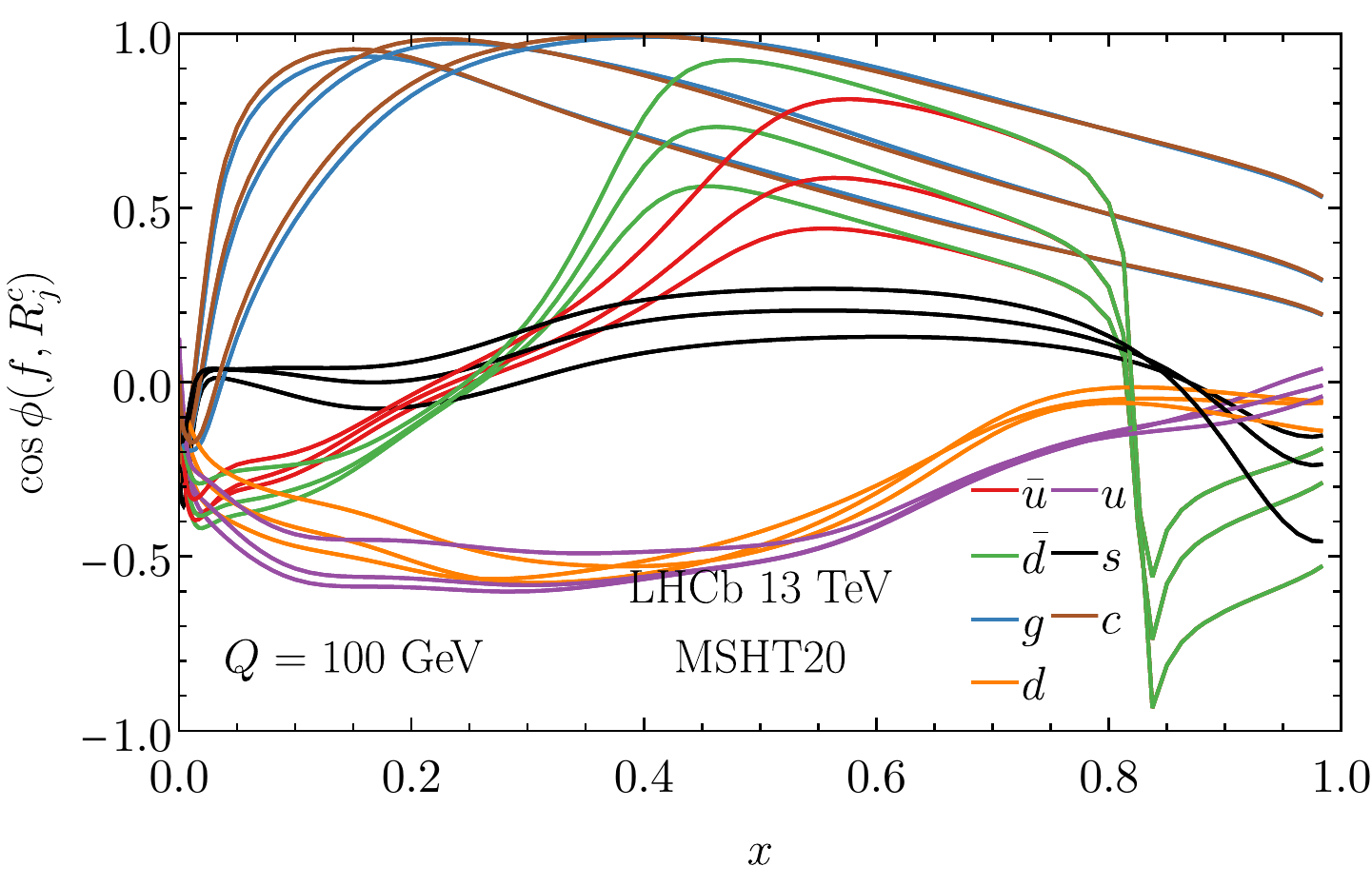}
    \includegraphics[width=0.32\textwidth]{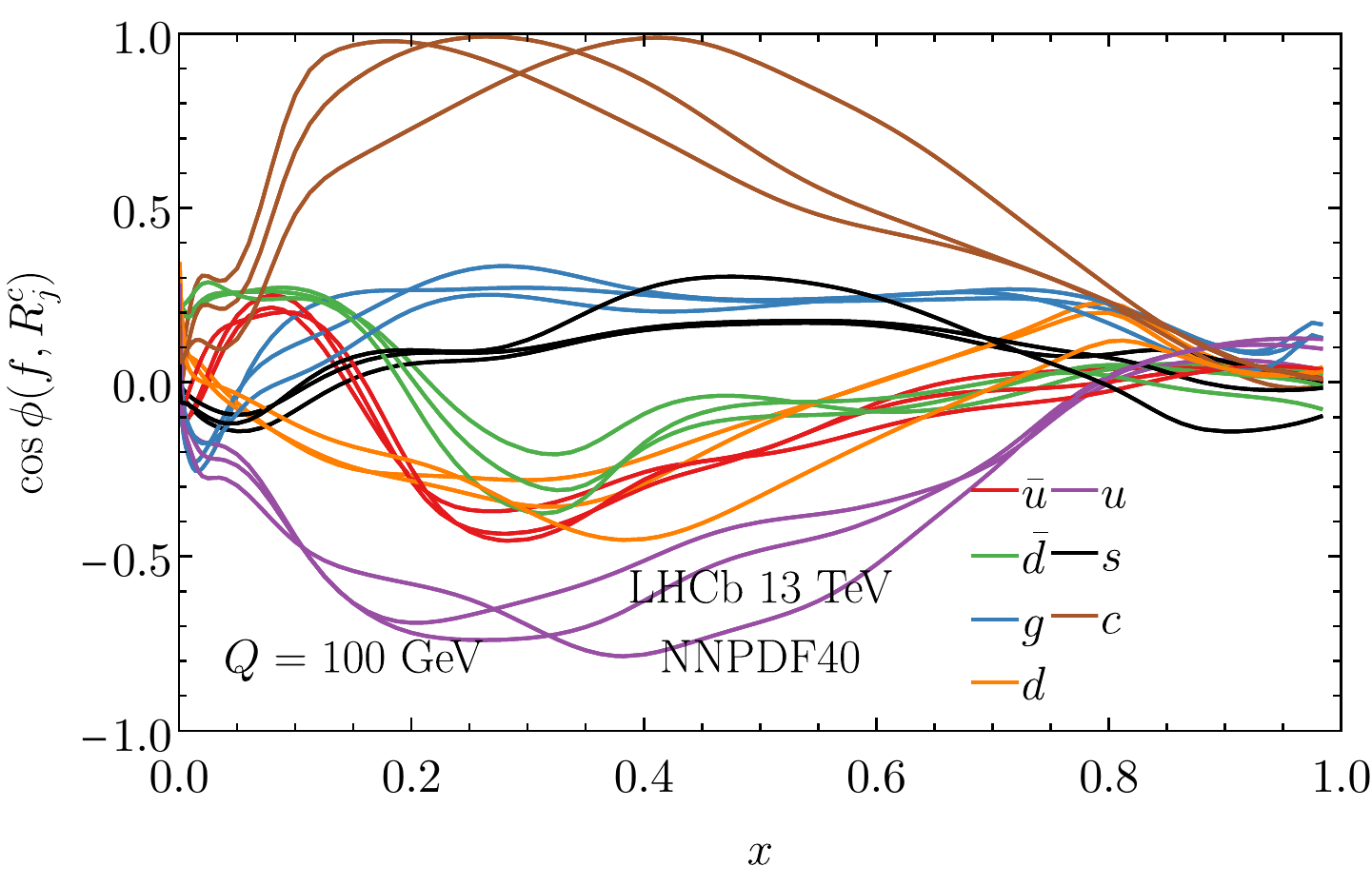}    
    \includegraphics[width=0.32\textwidth]{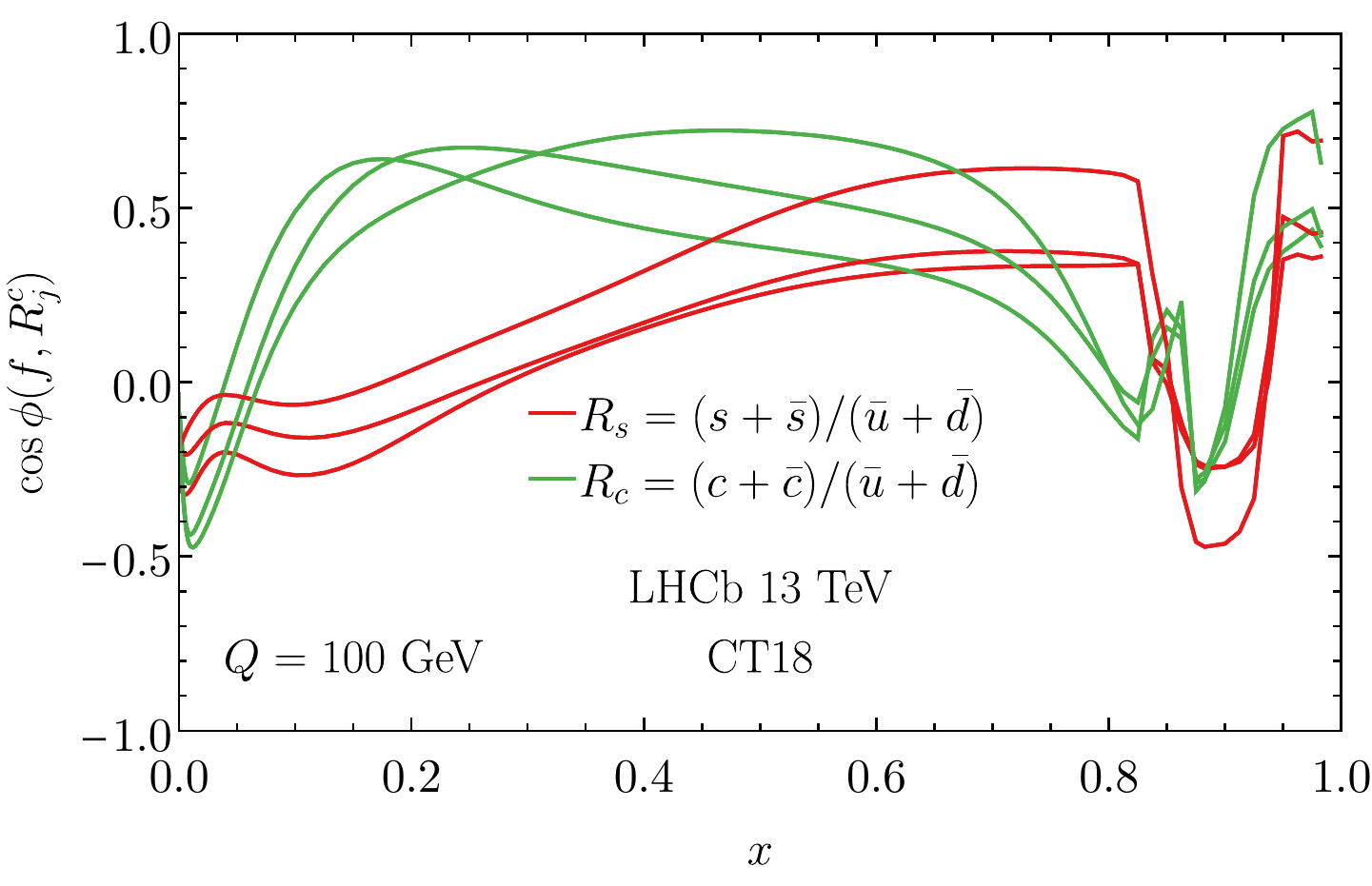}
    \includegraphics[width=0.32\textwidth]{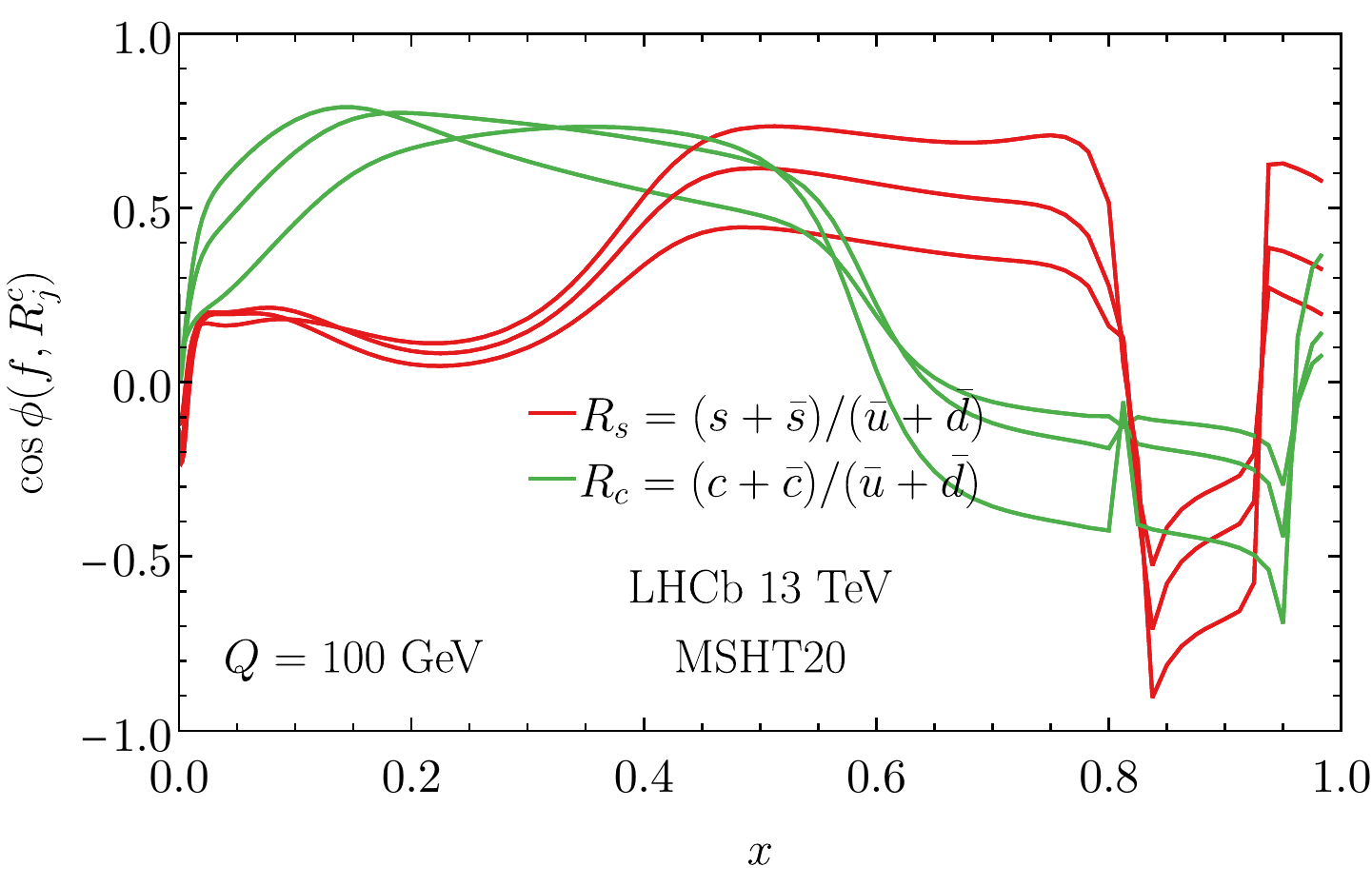}
    \includegraphics[width=0.32\textwidth]{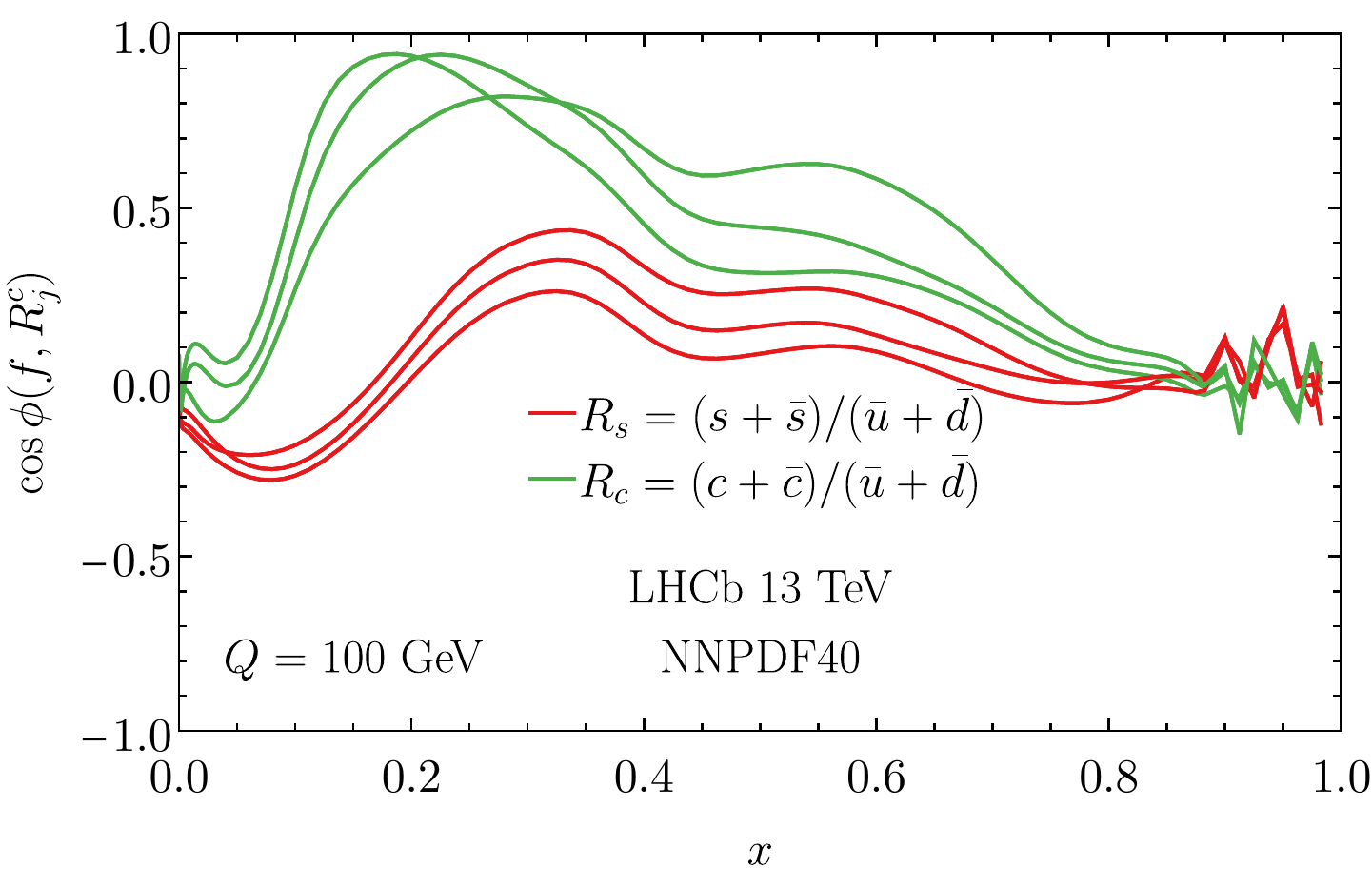}    \caption{The correlation between the CT18 (left column), MSHT20 (center column), and NNDF4.0 (right column) PDFs at $Q=100~\GeV$ and the LHCb 13 TeV cross sections $\sigma(Zc)$ (first row), $\sigma(Zj)$ (second row), as well as the corresponding $R_j^c$ ratios (third and fourth rows). The three curves of the same color correspond to the three LHCb rapidity bins~\cite{LHCb:2021stx}, where the respective $Z$ rapidities increase from left to right.}
    \label{fig:Corr}
\end{figure}

In Fig.~\ref{fig:Corr}, we plot the correlation angles between the 13 TeV LHCb $Z+c$ and $Z+j$ cross-section ratio, $R_j^c=\sigma(Zc)/\sigma(Zj)$, and the NNLO PDFs of CT18~\cite{Hou:2019efy}, MSHT20~\cite{Bailey:2020ooq}, and NNPDF4.0~\cite{NNPDF:2021njg}. The correlation angles for Hessian error sets and Monte-Carlo replicas are defined in Ref.~\cite{Gao:2013bia}.
As we see, $Z+c$ production is strongly correlated with the charm PDF, which is reflected in all three groups. In the CT18 and MSHT20 global analyses, the charm is perturbatively generated through gluon splitting, $g\to c\bar{c}$. As a consequence, correlations with the charm PDF closely follow those of the gluon. However, this pattern does not hold for NNPDF, since the freely-fitted charm PDF they include by default is no longer directly dependent on the gluon distribution. Meanwhile, we also find sizable correlations between the $Z+c$ data and the light-quark sea PDFs in CT18 and MSHT20, which originates from the momentum sum rule. Upon closer inspection, we see that the strange PDF correlations in CT18 roughly follow those of $\bar{u},\bar{d}$, whereas the MSHT $s$-PDF shows less correlation, presumably due to the additional parameters introduced for nucleon strangeness in the MSHT20 fit~\cite{Bailey:2020ooq}. Compared to $Z+c$, we see in the second row of Fig.~\ref{fig:Corr} that the $Z+j$ cross section is most strongly (anti-)correlated with the gluon PDF, particularly in the low-$x$ region. We also point out that the correlations with the sea-quark PDFs exhibit a sharp turnover near $x\gtrsim0.8$ for both CT18 and MSHT20; this feature is attributable to a sign-change in the extrapolated region.

In the third row of Fig.~\ref{fig:Corr}, we plot the usual PDF correlations with the measured cross-section ratio, $R_j^c=\sigma(Zc)/\sigma(Zj)$; complementary to this, the fourth and final row gives the analogous correlations between the LHCb $R_j^c$ ratio data and the strange- and charm-suppression ratios, $R_{s(c)}\equiv [s(c)+\bar{s}(\bar{c})]/[\bar{u}+\bar{d}]$. The behavior
of the $R_j^c$ correlations is mainly induced by the $Z+c$ absolute cross section, as shown in the corresponding panels of the first row.

In conclusion, we observe strong correlations between the LHCb $\sigma(Zc)/\sigma(Zj)$ measurements and the charm PDF, which suggest this measurement may potentially shed some light on FC, as discussed in the main Letter. However, we also find that the correlations between the various PDF flavors depend on the various parametrizations and methodologies adopted by different PDF-fitting groups. To account for these differences, theory
developments at NNLO, with parton shower effects included as discussed in the main Letter, will be essential for ensuring a clean interpretation of future $Z+c$ results.

\subsection*{\bf Alternative fits with non-vanishing strangeness and charm asymmetries}
\begin{figure}[htb]
\center
\includegraphics[width=0.32\textwidth]{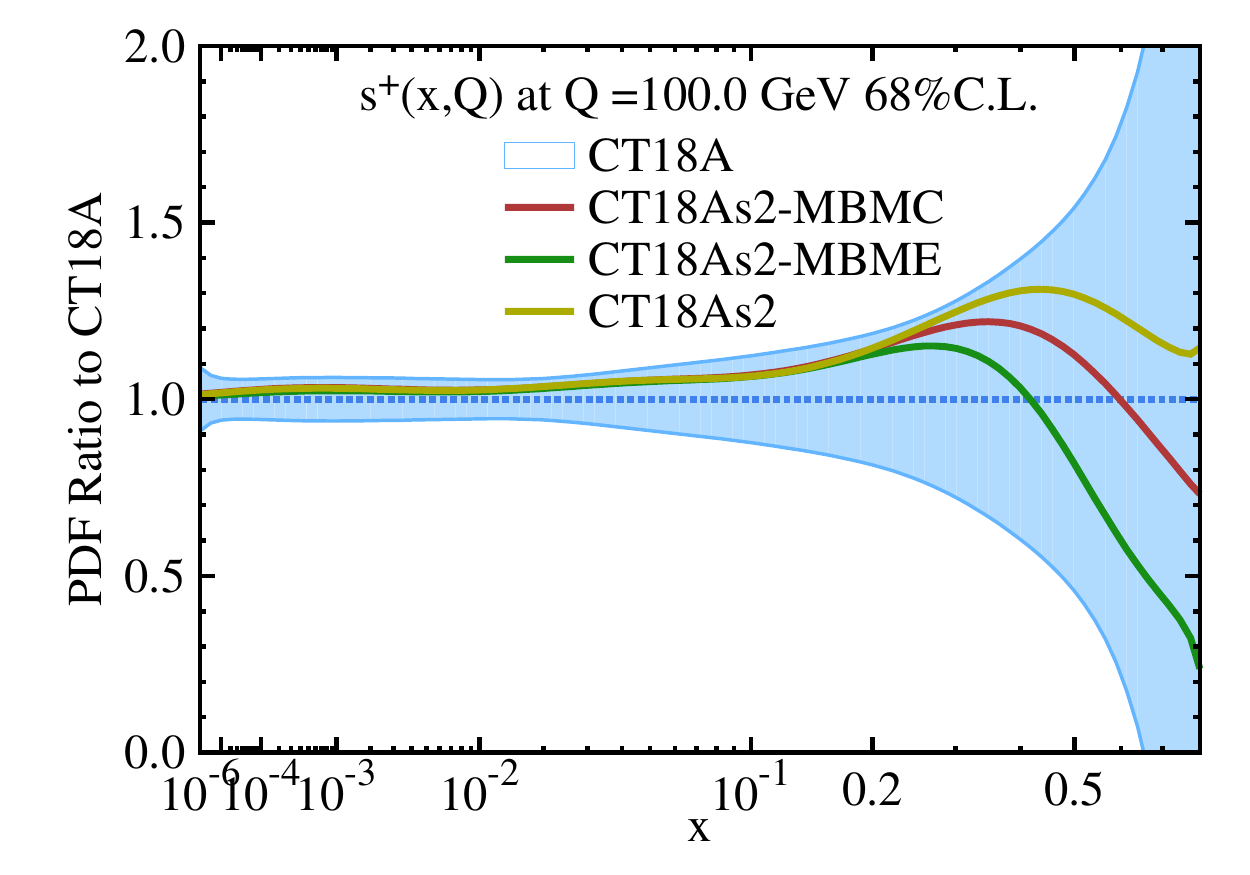}
\includegraphics[width=0.32\textwidth]{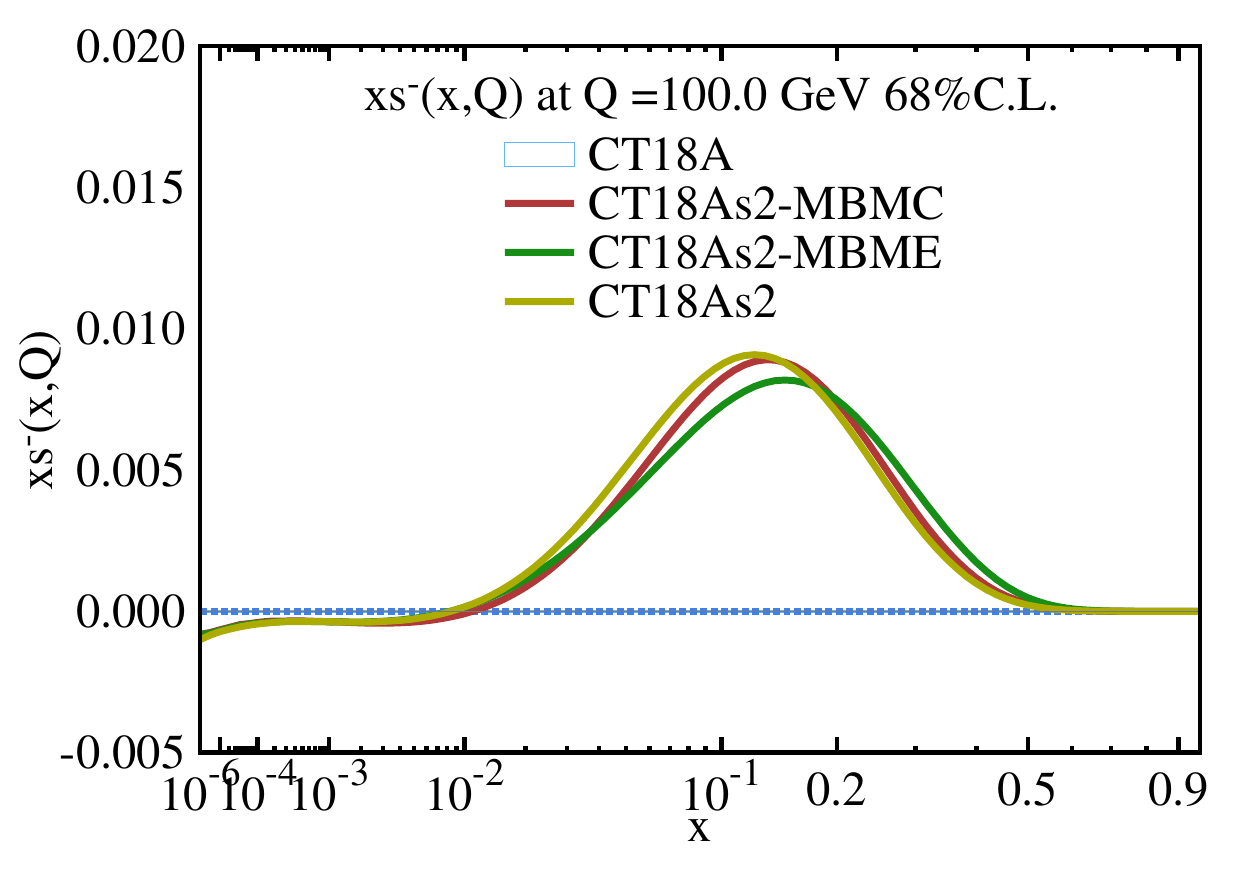}
\includegraphics[width=0.32\textwidth]{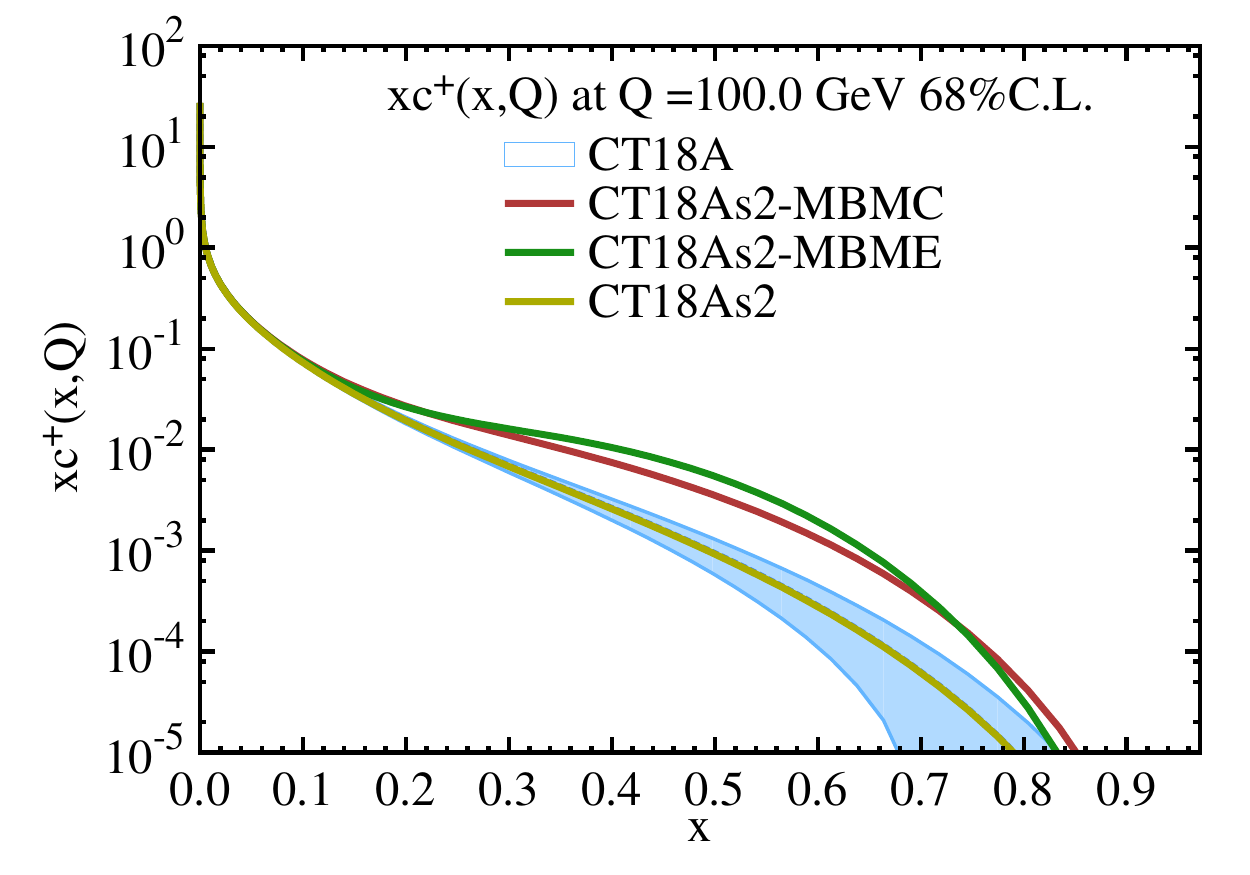}
\caption{Comparison for various  parametrizations with $s \neq {\bar s}$ and $c \neq {\bar c}$ at the $Q_0$ scale.}
	\label{fig:CT18As_IC}
\end{figure}

In Fig.~\ref{fig:CT18As_IC}, we compare a number of additional fit variants to illustrate the interplay between fitted strange and FC mentioned in the main text. Of these, CT18A assumes $s\! =\! {\bar s}$ at the $Q_0$ scale, while all others allow  $s\! \neq\! {\bar s}$. Moreover, both CT18A and CT18As2~\cite{Hou:2022sdf} assume $c\!=\!{\bar c=0}$ at the $Q_0$ scale, while the other two fits (MBMC and MBME) involve non-vanishing strangeness {\it and} charm asymmetries.
In particular, these plots demonstrate that the inclusion of FC according to the MBM with $c\!\neq\!\bar{c}$ leads to a modest suppression of $s^+(x)$ at very high $x$ relative to the CT18As2 fit which allowed $s\!\neq\!\bar{s}$. Meanwhile, the strange asymmetry is effectively unchanged by the incorporation of FC with a charm-anticharm asymmetry.

\section*{Additional plots of PDFs}
\begin{figure}[htb!]
\center
\includegraphics[width=0.47\textwidth]{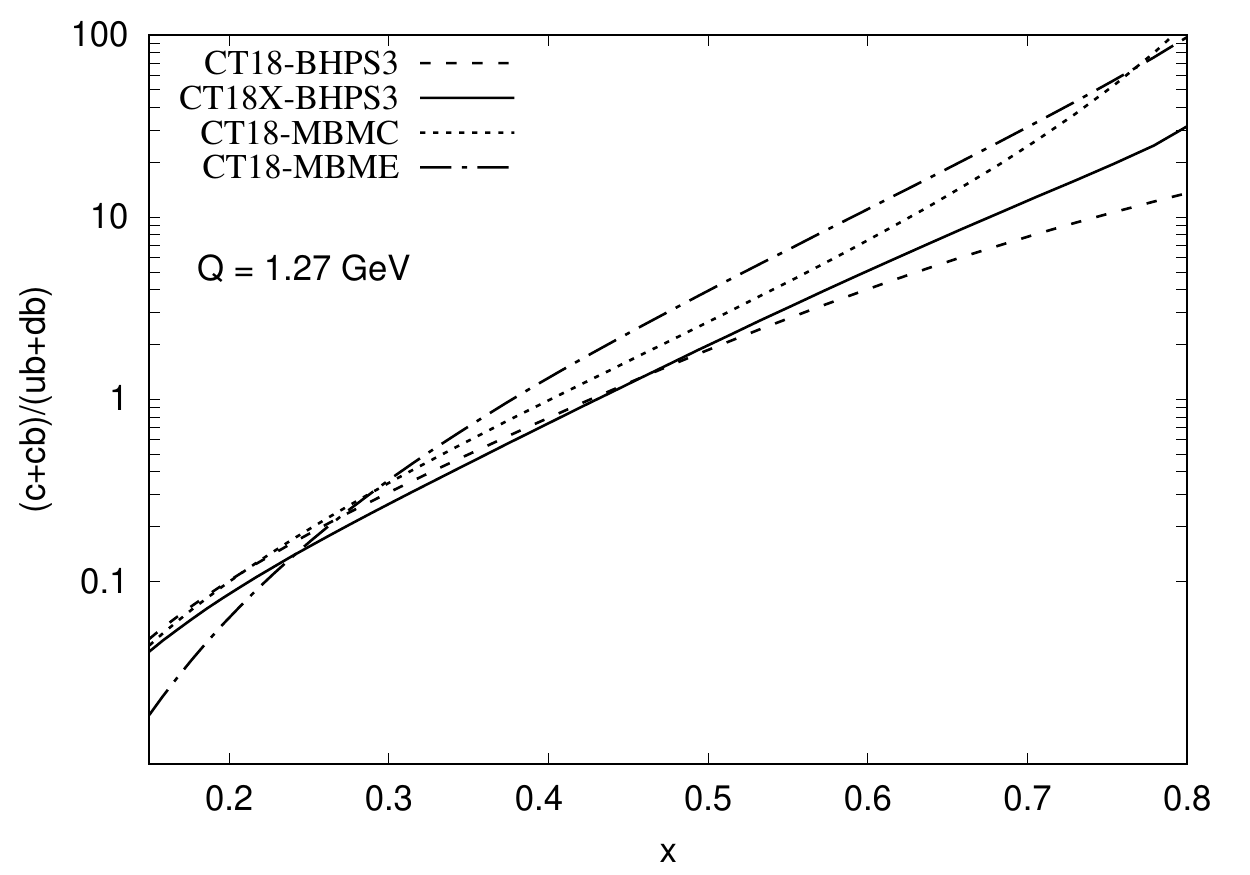}
\includegraphics[width=0.47\textwidth]{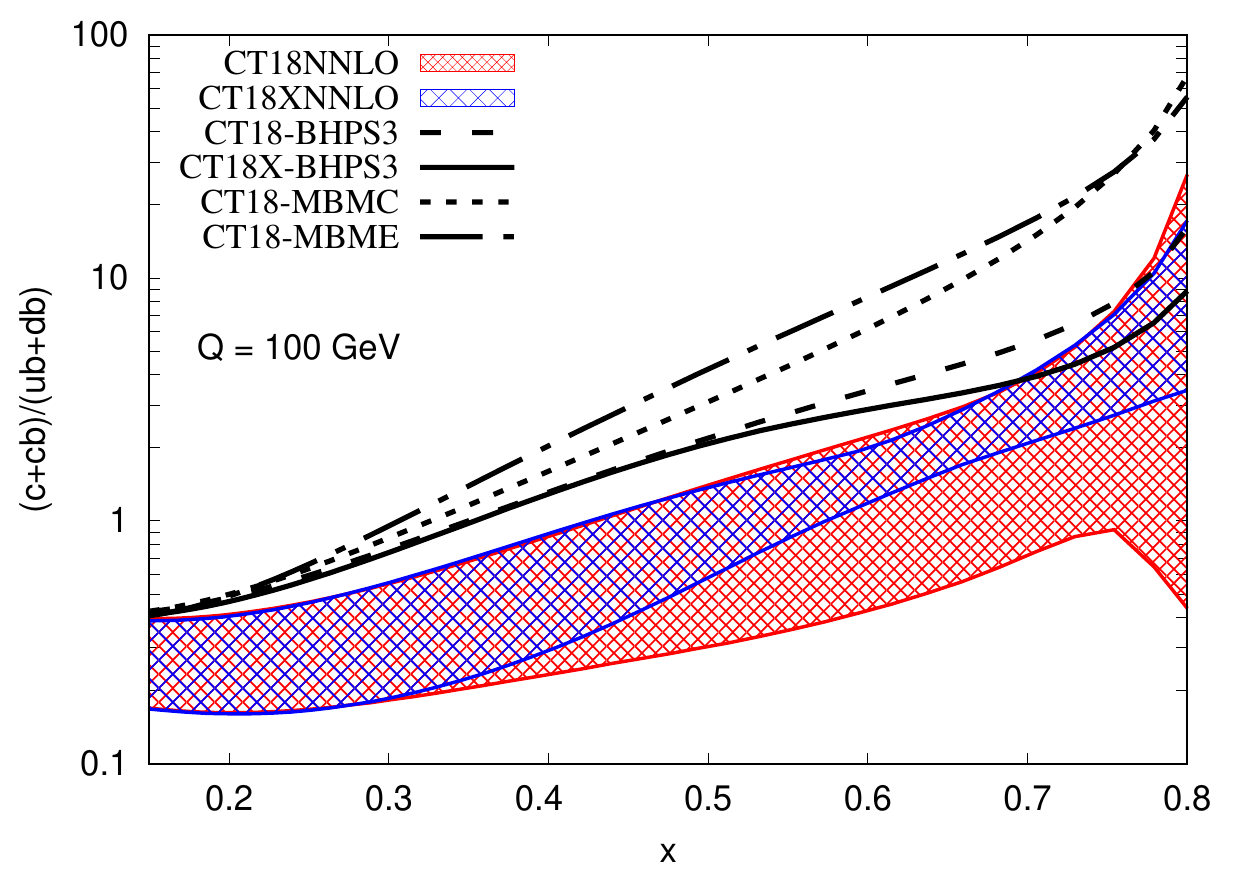}
	\caption{$R_c\! =\!(c+\bar{c})/(\bar{u}+\bar{d})$ ratio at $Q_0=1.27$ GeV and $Q=100$ GeV.  
\label{fig:Rc-ratio}}
\end{figure}

For completeness, we include several more figures further illustrating the behavior of the CT18 FC PDFs. Figure~\ref{fig:Rc-ratio} compares the charm-suppression ratio $R_c(x,Q)\equiv c^+ (x,Q)/\left(\bar u(x,Q) + \bar d(x,Q)\right) $ in the CT18 NNLO and CT18 FC models at $Q=1.27$ and $100$ GeV.
These plots suggest that, given sufficient magnitude, FC can
enhance the total charm PDF relative to the light-quark sea in a fashion that persists to high scales in the high-$x$ region.

\begin{figure}[htb!]
\center
\includegraphics[width=0.47\textwidth]{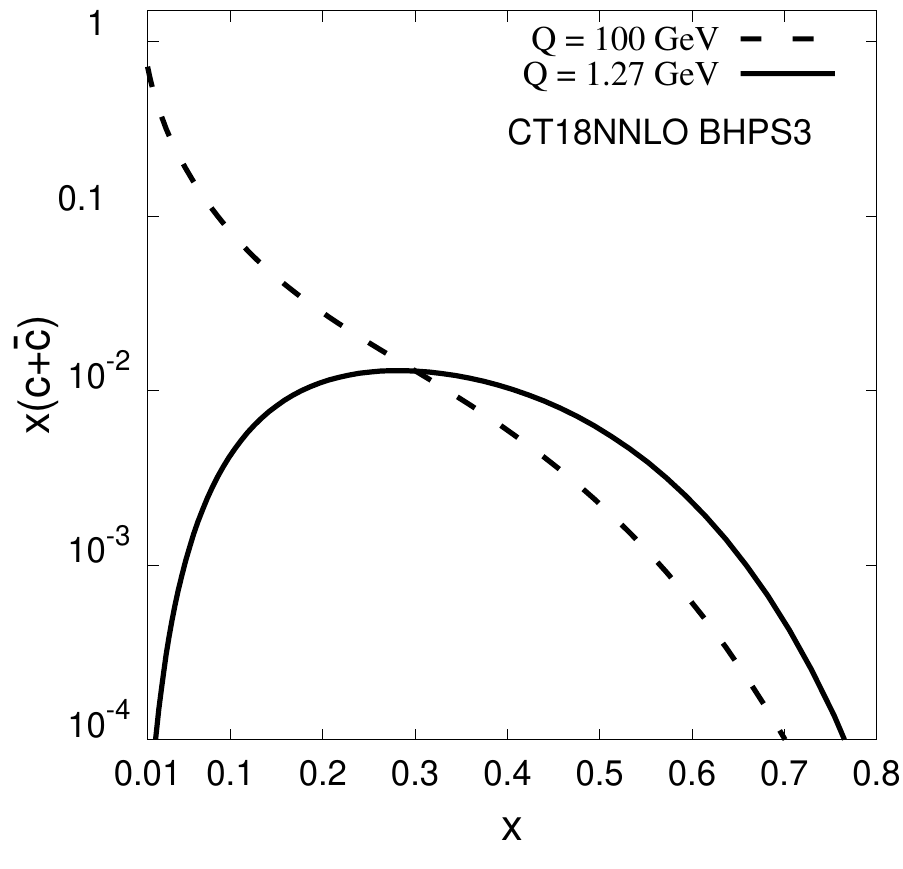}
\caption{Impact of DGLAP evolution from $Q=1.27$ GeV to $Q=100$ GeV on the $x(c+\bar{c})$ PDF combination for the BHPS3 model obtained within the CT18NNLO global analysis.   
\label{fig:bhps3-evol}}
\end{figure}

Figure~\ref{fig:bhps3-evol} illustrates DGLAP evolution of $xc^+ (x,Q)$ in the BHPS3 model between the scales $Q=1.27$ and 100 GeV. Our result for $xc^+ (x,Q)$ at $Q_0 =1.27$ GeV reflects in part the rate of backward DGLAP evolution from higher scales $Q > Q_0$ where the data are fitted. As such, backward DGLAP evolution tends to  increase (suppress) $xc^+ (x,Q)$ at $x \gtrsim 0.1$ ($x \lesssim 0.1$). Given that IC is an NNLO effect, control of higher-order uncertainties beyond NNLO in DGLAP evolution is necessary for accurate determination of $xc^+ (x,Q)$ at $Q_0$.


\end{document}